\documentclass{aa}

\usepackage{amsmath}
\usepackage{graphicx}
\usepackage{overpic}
\usepackage{xcolor}

\bibpunct{(}{)}{;}{a}{}{,}

\newcommand{\corr}[1]{#1}

\begin{document}
 
\title{fastRESOLVE: fast Bayesian imaging for aperture synthesis
in radio astronomy}

\author{M.~Greiner\thanks{\email{maksim@mpa-garching.mpg.de}}\inst{\ref{inst:MPA}}
  \and{V.~Vacca}\inst{\ref{inst:MPA}}\inst{\ref{inst:INAF}}
  \and{H.~Junklewitz}\inst{\ref{inst:AIFA}}
  \and{T.~A.~En{\ss}lin}\inst{\ref{inst:MPA}}
  }

\institute{Max-Planck-Institut f\"ur Astrophysik, Karl-Schwarzschild-Str.~1, 85748 Garching, Germany \label{inst:MPA}
  \and Argelander-Institut f\"ur Astronomie,  Auf dem H\"ugel 71, 53121 Bonn, Germany \label{inst:AIFA}
  \and INAF - Osservatorio Astronomico di Cagliari, Via della Scienza 5, 09047 Selargius, Italy \label{inst:INAF}}

\date{Received DD MMM. YYYY / Accepted DD MMM. YYYY}

\abstract{
The standard imaging algorithm for interferometric radio data, CLEAN, is optimal for point source observations, but suboptimal for diffuse emission. 
Recently, \textsc{Resolve}, a new Bayesian algorithm has been developed, which is ideal for extended source imaging. Unfortunately, \textsc{Resolve} is computationally very expensive.
In this paper we present fast\textsc{Resolve}, a modification of \textsc{Resolve} based on an
approximation of the interferometric likelihood that allows us to avoid
expensive gridding routines and consequently gain a factor of roughly 100
in computation time. Furthermore, we include
a Bayesian estimation of the measurement uncertainty of the visibilities into the imaging, a procedure not applied in aperture synthesis before. The
algorithm requires little to no user input compared to the standard method CLEAN
while being superior for extended and faint emission. We apply the
algorithm to VLA data of Abell 2199 and show that it resolves more detailed structures.
}


\titlerunning{fast\textsc{Resolve}}
\authorrunning{M.~Greiner et al.}
\maketitle

\section{Introduction}

A fundamental observation method in modern radio astronomy is the aperture synthesis technique 
\citep[see~e.g.][]{Ryle-1960, Thompson-1986, Finley-2000}.
Antennas of large interferometers are correlated to achieve resolutions comparable to a single dish instrument of the size of the whole array.
The downside of the technique is an increase in the complexity of data processing, since the interferometer only measures irregularly spaced sample points of the Fourier transform of the sky brightness leaving large unsampled regions in the Fourier plane. An inverse Fourier transform would suffer from strong aliasing effects severly distorting the image and misplacing regions of high brightness etc. Therefore, the brightness distribution on the sky has to be estimated in a more elaborate way.

The most widely used imaging algorithm in radio astronomy is CLEAN, developed by \cite{Hoegbom-1974}. The underlying assumption of CLEAN is that the image is composed of uncorrelated point sources. Consequently, CLEAN is very effective in imaging observations of point source dominated fields 
\citep[see~e.g.][]{Thompson-1986, Taylor-1999, Sault-2007}.
There are multiple variants and extensions to CLEAN reducing the computational effort \citep{Clark-1980}, improving the performance for multi-frequency observations \citep{Sault-1994}, and implementing corrections \citep{Schwab-1984}.
Reconstruction of extended objects has been adressed by implementing differently scaled kernel functions instead of sharp point sources as a basis, resulting in the Multi-Scale (MS-)CLEAN and Adaptive Scale Pixel (ASP) algorithm \citep{Bhatnagar-2004, Cornwell-2008, Rau-2011}. However, how to choose the scales for MS-CLEAN remains a non-trivial task left to the user and an implementation of the ASP algorithm is yet to be published.

The underlying assumption of the image being composed of point sources (or kernel functions) remains a fundamental ingredient in CLEAN, which hinders its performance in reconstructing extended diffuse sources. Furthermore, it is not known how to propagate measurement uncertainty through CLEAN and consequently, no uncertainty map for the reconstruction is provided.
To overcome these issues a new algorithm called \textsc{Resolve} (\textbf{R}adio \textbf{E}xtended \textbf{SO}urces \textbf{L}ognormal decon\textbf{V}olution \textbf{E}stimator) has been developed by \cite{Junklewitz-2016}. \textsc{Resolve} was developed within the framework of \textit{information field theory} of \cite{Ensslin-2009} and fulfills two main objectives\footnote{The objectives are a direct quote from \cite{Junklewitz-2016}.}:
 \begin{enumerate}
  \item It should be optimal for extended and diffuse radio
sources.
\item It should include reliable uncertainty propagation and
provide an error estimate together with an image reconstruction.
 \end{enumerate}
\textsc{Resolve} is a Bayesian algorithm. As such it needs to transform between image and data space many times in order to calculate the estimate of the brightness distribution on the sky. These transformations are costly, as they involve a Fourier transform to an irregular grid. In consequence, \textsc{Resolve} is computationally much slower than CLEAN. However, the Bayesian approach has the advantage that it requires fewer input parameters by the user since regularization of the result happens automatically and the additional time needed to determine some of these input parameters in CLEAN partly compensates for the additional compuation time.
Nevertheless, significant performance gains have to be achieved in order to make \textsc{Resolve} a widely used imaging algorithm for interferometric data. Furthermore, a Bayesian algorithm finds an estimate by weighing the prior against the likelihood. This weighting relies on an accurate description of the uncertainty of the recorded visibilities which is not always given for interferometric data. Algorithms like CLEAN, which operate solely on the likelihood are insensitive to a global factor in the measurement uncertainty, but they require a manually chosen cut-off criterion to avoid overfitting. Choosing this criterion properly is again related to estimating the measurement uncertainty. 

In this paper we introduce fast\textsc{Resolve}, an algorithm that reduces the computational cost of \textsc{Resolve} by avoiding the costly irregular Fourier transforms. Instead, the data are gridded to a regular Fourier grid in an information theoretically optimal way\corr{, which is conceptually similar to gridded imaging.}
Additionally, we include a measurement variance estimator into the procedure allowing us to overcome the dependency of an accurate description of the measurement uncertainty. Furthermore, we introduce a hybrid approach to simultaneously reconstruct point-sources and diffuse emission using a preprocessing similar to CLEAN. Thus, we tackle the major disadvatages of \textsc{Resolve} compared to CLEAN, bringing it a step closer to a widely applicable imaging algorithm for interferometric data.

\corr{CLEAN and \textsc{Resolve} are of course not the only deconvolution algorithms in the context of radio astronomy. Most notably, there is the Maximum Entropy Method for imaging \citep[][]{Gull-1978, Cornwell-1985}.
Other approaches include non-negative-least-squares \citep[][]{Briggs-1995} and compressed sensing techniques \citep[][]{Wiaux-2009, Carrillo-2012}. For a further discussion of alternative methods in the context of \textsc{Resolve} we refer to \cite{Junklewitz-2016}.}

The remainder of this paper is structured as follows. In section \ref{sec:algorithm} we derive fast\textsc{Resolve} focusing on the approximations and add-ons that distinguish fast\textsc{Resolve} from \textsc{Resolve}. In section \ref{sec:application} fast\textsc{Resolve} is tested and compared to CLEAN using archival data of Abell 2199. Section \ref{sec:summary} provides a summary of the findings as well as an outlook to further developments.

\section{Algorithm}
\label{sec:algorithm}

\subsection{Signal model}

The signal of interest is the intensity of electromagnetic radiation at some frequency across a patch of the sky, $I(x,y)$.
 The position on the sky is here described by the Cartesian coordinates $x$ and $y$. The coordinates span a plane tangential to the celestial sphere with the point of tangency being the center of the observation. This plane is called the $uv$-plane.
 The relationship between the intensity $I$ on the sky and the visibilities $V$ an interferometer records is described by the interferometer equation, which -- for a small patch of the sky \citep[see~e.g.][]{Thompson-1986} -- is essentially a Fourier transformation:
 \begin{equation}
  V(u,v) = \int\!\!\mathrm{d}x\mathrm{d}y\, e^{-2\pi i\, (u x + v y)} B(x,y) \, I(x,y).
  \label{eq:interferometer}
 \end{equation}
The term $B(x,y)$ describes the primary beam of the instrument. The coordinates $u$ and $v$ describe the displacement (projected onto the $uv$-plane)  between a pair of antennas in units of the observed wavelength.
Since there is a finite set of antenna combinations forming a finite set of $uv$-locations only discrete visibilities are observed.
We label these with an index $i$,
\begin{equation}
 V_i \equiv V(u_i,v_i)
\end{equation}

For notational convenience we introduce the response operator $R$ describing the linear relationship between the data $d$ and the intensity $I$,
\begin{equation}
R_i(x,y) = \exp\left( -2\pi i\, (u_i x + v_i y)\right) B(x,y). 
\end{equation}
To further improve readability we use a compact matrix notation for objects living in the continuous sky.
The application of $R$ to $I$ is denoted as,
\begin{equation}
 V_i = (RI)_i := \int\!\!\mathrm{d}x\mathrm{d}y\, R_i(x,y) I(x,y).
\end{equation}
Similarly we define scalar products of continuous quantities,
\begin{equation}
 I^\dagger j := \int\!\!\mathrm{d}x\mathrm{d}y\, I^*(x,y)\, j(x,y),
\end{equation}
where ${}^*$ denotes complex conjugation. The $\dagger$ symbol therefore denotes transpositions and complex conjugation. For objects living in the discrete data space, there are analogous operations with sums instead of integrals, e.g.,
\begin{equation}
 (R^\dagger V)(x,y) := \sum\limits_i R_i^*(x,y)V_i.
\end{equation}

The incompleteness of the $uv$-plane is the main challenge in interferometric imaging.
Additionally, the recorded data are subject to instrumental noise. If the full $uv$-plane was measured and there was no instrumental noise, the intensity $I$ could be simply recovered by inverse Fourier transformation. The noise is mostly of thermal origin and here assumed to be signal independent\footnote{The independence of the noise is of course an approximation.}, additive Gaussian noise,
\begin{equation}
\begin{split}
 d & = V + n, \\
 \mathcal{P}(n) & = \mathcal{G}(n,N),
\end{split}
\end{equation}
where $\mathcal{G}$ describes a multivariate Gaussian probability distribution,
\begin{equation}
\mathcal{G}(n,N) := \left| 2 \pi N \right|^{-\frac{1}{2}} \exp\!\left( -\frac{1}{2} n^\dagger N^{-1} n \right),
\end{equation}
with the covariance matrix $N := \left\langle n n^\dagger \right\rangle_{\mathcal{P}(n)}$. The covariance matrix is assumed to be diagonal in this paper, meaning that the noise contamination of different visibilities is not correlated,
\begin{equation}
 N_{ij} = \delta_{ij} \sigma_i^2.
 \label{eq:noise_variance}
\end{equation}
Here, $\sigma_i$ is the $1\sigma$ uncertainty interval of the data point $i$.

\subsection{Irregular sampling}

Any algorithm that aims to find an estimate of $I$ given $d$ will have to apply the response $R$ and thus evaluate the interferometer equation Eq.~\eqref{eq:interferometer} multiple times in order to ensure compatibility between the data and the estimated intensity. Therefore, the speed of this operation is of crucial importance for the performance of the algorithm and decisive in whether the algorithm is feasible to apply or not.
The interferometer equation Eq.~\eqref{eq:interferometer} involves a Fourier transform. In numerical applications this is typically carried out using a Fast Fourier transform (FFT). The FFT scales as $\mathcal{O}(N \log N)$ which is much faster that the $\mathcal{O}(N^2)$ of a direct Fourier transform. However, the FFT is only applicable if the sampling points are arranged in a regular grid. 
For a two-dimensional space such as a patch of the sky this means that $x$ and $y$ have to be distributed according to
\begin{equation}
 x = j \Delta x,\quad\text{for}\ -N_x/2\leq j < N_x/2,
\end{equation}
where $j$ is an integer and $N_x$ is the number of pixels in the $x$-dimension and $\Delta x$ is the pixel edge length\footnote{The origin of the coordinate system can of course be shifted arbitrarily.}. With a corresponding equation for $y$ and the assumption of periodic boundary conditions the sampling theorem states that a conjugate basis of $k$ and $q$ allows for lossless back and forth Fourier transformations if $k$ and $q$ are distributed according to
\begin{equation}
\begin{split}
 k & = j \Delta k,\quad\text{for}\  -N_x/2\leq j < N_x/2,\\
 \Delta k & = \frac{1}{N_x\Delta x}
\end{split}
\label{eq:k-spacing}
\end{equation}
with a corresponding expression for $q$ and $y$. Only with such a correspondence between the sample points of $x$, $y$ and $k$, $q$ is the FFT applicable.

The sample points of $x$ and $y$ can be freely chosen in the interferometer equation. However, the sample points on the left-hand-side are the $uv$-points sampled by the interferometer and cannot be freely chosen. There is basically no antenna setup in the real world where the $uv$-points can be brought into a form like Eq.~\eqref{eq:k-spacing} with an acceptable number for $N_x$. Simply mapping the $uv$-points to their closest (regular) $kq$-point yields strong aliasing effects contaminating the results. Calculating the aliasing explicitly yields a gridding operation between the visibility in a regular Fourier grid and the visibilities at irregularly spaced $uv$-points:
\begin{equation}
\begin{split}
 V(k,q) = \sum_j G_j(k,q) V_j,
\end{split}
\end{equation}
where $G$ is the gridding operator (see Appendix \ref{sec:aliasing}),
\begin{equation}
 G_j(k,q) \approx \frac{1}{\Delta k\, \Delta q} \mathrm{sinc}\!\left( \frac{k-u_j}{\Delta k} \right) \mathrm{sinc}\!\left( \frac{q-v_j}{\Delta q} \right).
\end{equation}
In this paper we use the normalized $\mathrm{sinc}$ function, \mbox{$\mathrm{sinc}(x) := \sin(\pi x)/(\pi x)$} (with \mbox{$\mathrm{sinc}(0) = 1$} of course). To make the notation more compact, we will use the vector notation $\vec{k} \equiv (k,q)$ and $\vec{x} \equiv (x,y)$ throughout the rest of this paper.

\subsection{The likelihood}
\label{sec:Mdiag}

The likelihood is a probability density function describing the possible outcomes of a measurement assuming a known signal configuation. Since we are assuming signal independent and additive Gaussian noise, the likelihood is a Gaussian centered on the noise-free visibilities,
 \begin{equation}
 \begin{split}
  \mathcal{P}(d|I) & = \mathcal{G}(d - RI, N)\\
  & = \left| 2 \pi N \right|^{-\frac{1}{2}} \exp\left( -\frac{1}{2} \left(RI - d\right)^\dagger N^{-1}\left(RI - d\right) \right).
  \end{split}
  \label{eq:likelihood}
 \end{equation}
The goal of our inference is to find an estimate of the intensity $I$. Therefore, the only terms of the likelihood that are of relevance to us are the ones dependent on $I$,
\begin{equation}
\mathcal{P}(d|I) \propto \exp\!\left( -\frac{1}{2} I^\dagger M I + j^\dagger I \right),
\end{equation}
with the measurement precision operator $M$ and the information source $j$,
\begin{equation}
 \begin{split}
  M & = R^\dagger N^{-1} R, \\
  j & = R^\dagger N^{-1} d.
 \end{split}
\end{equation}

 The measurement precision $M$ can be represented as
 \begin{equation}
  M = \hat{B} F \hat{B},
 \end{equation}
 where $\hat{B}$ is a diagonal operator in position space with the primary beam on the diagonal and
 \begin{equation}
 \begin{split}
  F(\vec{k}, \vec{k}') & = \sum\limits_{j,j'} G^*_j(\vec{k}) \left(N^{-1}\right)_{jj'} G_{j'}(\vec{k}')\\
                & = \sum\limits_j \frac{1}{\sigma_j^2} \,G^*_j(\vec{k})\, G_{j}(\vec{k}').
 \end{split}
 \end{equation}
If the sampled $uv$-points were matching the $k$ and $q$ points of the regular Fourier grid perfectly, the operator $F$ would be a simple (inverse noise weighted) mask in Fourier space, a diagonal operator. The points do not match, of course, so $F$ is not diagonal. The full matrix $F$ cannot be stored in the memory of a computer in realistic scenarios, so to apply $F$ the gridding operator $G$ has to be applied twice, which is numerically expensive, even though it can be done quicker than by invoking a direct Fourier transform \citep[see e.g.][]{Briggs-1999}.

We want to approximate $F$ by $\tilde{F}$, a diagonal operator in Fourier space, while preserving as much information as possible. This would enable us to apply $M$ without the expensive gridding operation, thus increasing the numerical speed significantly.
As a criterion of information loss we use the Kullback-Leibler divergence, sometimes called relative entropy,
\begin{equation}
 D_{\mathrm{KL}}\!\!\left[ \tilde{\mathcal{P}}(x) | \mathcal{P}(x) \right] := \int\!\!\mathcal{D}x\ \tilde{\mathcal{P}}(x) \ln \frac{\tilde{\mathcal{P}}(x)}{\mathcal{P}(x)},
\end{equation}
where $\mathcal{D} x \equiv \mathrm{d}x_1\mathrm{d}x_2\mathrm{d}x_3...$ denotes integration over the full phase space (all degrees of freedom of $x$).
By minimizing the Kullback-Leibler divergence with respect to $\tilde{F}$ we find that most information is preserved if $\tilde{F}$ is simply the diagonal of $F$,
\begin{equation}
 \tilde{F}(\vec{k}, \vec{k}') = \delta(\vec{k}-\vec{k}')\,F(\vec{k},\vec{k}).
\end{equation}
For the detailed calculation we refer the reader to Appendix \ref{sec:DKL_minimization}.
\corr{In Appendix \ref{sec:relation_to_standard} we discuss the relation of this approximation to standard practices like gridded imaging and major-minor-cycles.}

\subsection{Inference}

In this section we outline the inference algorithm. The algorithm is the original \textsc{Resolve} algorithm which is derived in detail by \cite{Junklewitz-2016} save for two additions (Sec.~\ref{sec:variance-estimation} and Sec.~\ref{sec:point-sources}) and an additional approximation (Appendix~\ref{sec:approximate-propagator}).
The inference combines the interferometric likelihood with a log-normal prior with a homogeneous and isotropic covariance,
\begin{equation}
\begin{split}
 &I  = \exp(s),\quad 
 \mathcal{P}(s)  = \mathcal{G}(s,S),\quad
 S  = \sum_i p_i S^{(i)},
 \label{eq:prior}
\end{split}
\end{equation}
where $p_i$ are unknown power spectrum parameters and $S^{(i)}$ are disjoint projection operators onto the different spectral bands\footnote{For a more detailed introduction of the concept we refer the reader to the references provided in this section.}. The reasoning behind this choice for a prior is that the intensity is a positive definite quantity. Furthermore, the algorithm should not assume any preferred direction or position prior to considering the data. This leaves us with an unknown power spectrum for which we choose a hyper-prior which prefers (but does not enforce) the power spectrum to be a power law \citep{Oppermann-2013, Ensslin-2011, Ensslin-2010}.
Using a maximum a posteriori Ansatz for $s$ and the empirical Bayes method\footnote{This can also be seen in the context of Variational Bayes, see e.g.~\cite{Selig-2015}.} \citep[see e.g.][]{Robbins-1956} for $p$ this yields the following set of equations for the logarithmic map $m$, its covariance estimate $D$, and the power spectrum $p$:
\begin{align}
% \begin{split}
 m& = \underset{s}{\mathrm{arg\,min}}\!\left( -\ln \mathcal{P}(s,d|p)\right),
 \label{eq:MAP-equations-m}\\
 D& = \left( \left.- \frac{\delta^2}{\delta s \delta s^\dagger} \ln \mathcal{P}(s,d|p) \right|_{s=m} \right)^{-1},
 \label{eq:MAP-equations-D}\\
 p_i& = \frac{q_i + \frac{1}{2}\mathrm{tr}\!\left[ \left( mm^\dagger + D \right) S^{(i)} \right]}{\alpha_i - 1 + \varrho_i/2 + \left( T \ln p \right)_i}.
 \label{eq:MAP-equations-p}
% \end{split}
\end{align}
Here, $\varrho_i=\mathrm{tr}\, S^{(i)}$ is the amount of degrees of freedom in the $i$th band whereas $q_i$, $\alpha_i$, and $T$ come from the hyper-prior of the power spectrum parameters (see \cite{Junklewitz-2016} \& \cite{Oppermann-2013} for details). In Eq.~\eqref{eq:MAP-equations-p} $D$ is used in an approximated form which is numerically cheaper to calculate. This approximation is described by Eq.~\eqref{eq:approx_propagator2} in Appendix~\ref{sec:approximate-propagator}. The solutions for $m$ and $D$ depend on $p$, which in turn depends on $m$ and $D$. Therefore, these equations have to be iterated until a fixed point is reached. The resulting map $m$ is the estimate of the logarithmic intensity $I$ while $D$ is an estimate of the posterior covariance of $m$.

\subsubsection{Measurement variance estimation}
\label{sec:variance-estimation}

The reconstruction of the diffuse emission is sensitive to the variances in the likelihood Eq.~\eqref{eq:likelihood}. However, these variances are often not available in a way that is consistent with Eq.~\eqref{eq:likelihood}. Therefore, we need to estimate these variances along with the emission itself.
The estimate for $\sigma^2$ is derived in Appendix~\ref{sec:variance-derivation} as
\begin{equation}
 \sigma^2_i = \left| \left(Re^m - d  \right)_i\right|^2.
 \label{eq:sigma-update}
\end{equation}
This equation is prone to overfitting, since individual data points can yield a very small $\sigma$ leading to an even stronger fit in the next iteration.
We therefore introduce a regularization,
\begin{equation}
 \sigma^2_i \leftarrow t\, \sigma^2_i + \frac{1-t}{N_\mathrm{data}} \sum_j \sigma^2_j,
\label{eq:sigma-regu}
\end{equation}
where $0\leq t \leq 1$. If the data are properly calibrated (e.g. outliers have been flagged), we recommend to set $t$ to 0, i.e., setting all $\sigma_i$ equal. If the data contain a significant amount of outliers $t$ has to be increased to allow the $\sigma_i$ to weigh them down. For such cases we found $t=0.9$ to be a good value.
If different data sets are combined we recommend to perform the regularization in Eq.~\eqref{eq:sigma-regu} separately for each data set.

\subsubsection{Point source removal}
\label{sec:point-sources}

The prior assumptions outlined in this section so far implicitly assume the intensity to be spatially correlated due to the prior assumption of a statistically homogeneous and isotropic log-normal field. This assumption is obviously not very well suited for point sources, which are present in most radio observations. Point sources are often orders of magnitude brighter than the extended sources of radio emission and can thus complicate the reconstruction of the latter, especially if the measurement operator in the likelihood has been approximated.
To overcome this we subtract point sources from the data by an iterative procedure somewhat similar to CLEAN.
To that end the intensity is divided into the extended emission intensity $I$ and the point source emission intensity $I_\mathrm{point}$. The goal is to determine $I_\mathrm{point}$ first and apply the algorithm for extended emission on the point source removed data. The iterative procedure is stopped when the amount of pixels containing a point source reaches a cut-off criterion.
One could of course use CLEAN itself for this, but we implemented our own procedure to understand all the subtleties. We describe it in Appendix~\ref{sec:point-source-details}.
A more rigorous procedure which combines \textsc{Resolve} with the D$^3$PO algorithm \citep{Selig-2015} is in preparation (Junklewitz et al.~in prep., working title point\textsc{Resolve}).

After $I_\mathrm{point}$ is determined and subtracted the remaining data should contain mostly extended emission and can be further processed using Eqs.~\eqref{eq:MAP-equations-m}~to~\eqref{eq:MAP-equations-p} to reconstruct the extended emission.

\subsubsection{Filter procedure}
\label{sec:filter-procedure}

The filter equations and procedures described in this section allow us to determine the point-like emission, diffuse emission, the power spectrum of the diffuse emission, and the noise level (measurement variance).
Due to the non-convex nature of the system of equations the starting point of the solving strategy is of importance.
The following initial guesses for the power spectrum and noise variances proved to be useful.

The initial power specturm is set to $p_i \propto \ell_i^{-3}$, where $\ell$ is the average length of the $k$-vectors in band $i$. The prefactor is chosen so that the variance of a corresponding random field is $\sim 1$.  The monopole $p_0$ is chosen to be practically infinity (but still a floating point number). This means that the relative fluctuations of $I$ are a priori assumed to be of order $e$ while its absolute scale is free.

The initial noise variances $\sigma_i^2$ are set to half of the average data power.
% $\sigma_i^2 = \frac{1}{2N_\mathrm{data}} \sum\limits_j \left| d_j \right|^2$.
This corresponds to an assumed signal-to-noise ratio of 1.

With these starting values we solve the equations in the following way:
\begin{enumerate}
 \item separate point sources according to Appendix~\ref{sec:point-source-details} \label{iter:point}
 \item update the power spectrum by iterating Eqs.~\eqref{eq:MAP-equations-m} to \eqref{eq:MAP-equations-p} several times
 \item repeat from step \ref{iter:point} until no more point sources are found
 \item update the variances according to Sec.~\ref{sec:variance-estimation} \label{iter:vari}
 \item update the map according to Eq.~\eqref{eq:MAP-equations-m} and repeat from step \ref{iter:vari} until the variances do not change anymore.
\end{enumerate}
The resulting estimates of $I=e^m$, $I_\mathrm{point}$, $p$, and $\sigma^2$ are our final results.

\subsubsection{Posterior uncertainty map}
\label{sec:uncertainty}

After the iterative procedure described in \ref{sec:filter-procedure} has converged, the posterior variance of $s$ can be approximated (within the maximum a posteriori approach) by
\begin{equation}
 \left\langle \left[s(\vec{x}) - \left\langle s(\vec{x}) \right\rangle_{\mathcal{P}(s|d)}\right]^2 \right\rangle_{\mathcal{P}(s|d)} \approx D(\vec{x},\vec{x}),
\end{equation}
where we use the approximative form of $D$ described Eq.~\eqref{eq:approx_propagator} in Appendix~\ref{sec:approximate-propagator}.
Using this result we estimate posterior variance of $I=e^s$ as
\begin{equation}
  \left\langle \left[I(\vec{x}) - \left\langle I(\vec{x}) \right\rangle_{\mathcal{P}(s|d)}\right]^2 \right\rangle_{\mathcal{P}(s|d)} \approx e^{m(\vec{x})}\left(e^{D(\vec{x},\vec{x})}-1\right)e^{m(\vec{x})}.
\end{equation}
It is important to note that this uncertainty estimate is derived within a saddle-point approximation to a non-convex problem. Consequently, it cannot detect artifacts in the maximum a posteriori solution $m$. The uncertainty estimate is rather to be interpreted as the uncertainty assuming $m$ was a good estimate. It does not consider that structures could be misplaced due to aliasing for example. However, it is still useful to have even a crude uncertainty estimate compared to having no estimate at all, which is the case for all established imaging methods for aperture synthesis.

Furthermore the uncertainty estimate only considers the statistical error of diffuse component of the image. It does not include an estimate of the systematic uncertainty. The point-like component is regarded as fixed. fast\textsc{Resolve} does not provide an uncertainty estimate for it. This will be fixed by the soon to be published point\textsc{Resolve} addition mentioned in Sec.~\ref{sec:point-sources}.

\section{Application}
\label{sec:application}

In this section we demonstrate the performance of fast\textsc{Resolve}. In the original publication \citep{Junklewitz-2016} of \textsc{Resolve} its performance on simulated data is tested and compared thoroughly to MS-CLEAN and the Maximum Entropy Method. Instead of reproducing these comparisons for fast\textsc{Resolve}, we validate the behavior of fast\textsc{Resolve} against standard \textsc{Resolve} to check that the approximations and add-ons in fast\textsc{Resolve} do not compromise the results.
These mock tests can be found in Appendix~\ref{sec:mocktest}. 
The tests show that fast\textsc{Resolve} is consistent with \textsc{Resolve}, but the latter recovers the input data to a higher accuracy. This is to be expected given the approximations involved.

Here we want to focus on the performance using real data. To this end we use observations of Abell 2199 and compare their fast\textsc{Resolve} images to their respective MS-CLEAN images.
This is the first time that a real data reconstruction using any variant of \textsc{Resolve} is published.

\subsection{Abell 2199}

We applied fast\textsc{Resolve} to radio observations of the cool core galaxy
cluster A2199. The brightest galaxy in this cluster is the powerful active galactic nucleus (AGN)
3C\,338, a restarting Fanaroff-Riley I radio galaxy with two symmetric
jets on parsec-scales, displaced from the large-scale emission of the
lobes. 3C\,338 is associated with the multiple nuclei optical cD galaxy
NGC\,6166, while X-ray observations of the cluster indicate the presence
of cavities associated with the lobes. This radio source has been studied
by several authors \citep[e.g.][]{Burns-1983, Fanti-1986, Feretti-1993, Giovannini-1998, Markevitch-2007}.
Recently, \cite{Vacca-2012} derived the intracluster magnetic field
power spectrum by using archival VLA observations over the frequency band
$1-8\,\mathrm{GHz}$. In this work we use the VLA data at $8145\,\mathrm{GHz}$ in C and BC
configurations presented in their paper.
The details of the observations are given in Table
\ref{tab:observations}. We refer to their paper for the description of the
data calibration. The combined dataset at $8145\,\mathrm{GHz}$ was used to produce
images of the source with the MS-CLEAN algorithm implemented in AIPS\footnote{
\url{http://www.aips.nrao.edu}
} with
natural, uniform, and intermediate weighting. \corr{MS-CLEAN was run with six scales: 0, 3, 5, 10, 20, and 40 arcsec.}
These images are compared with the image obtained with fast\textsc{Resolve}.

\begin{table*}[h]
\caption{Summary of the observations used in this paper. We refer to \cite{Vacca-2012} for further details.}
\label{tab:observations}
\begin{tabular}{cccccccc}
\hline\hline
Frequency& Bandwidth&Project& Date&Time& VLA configuration& RA (J2000) &
Dec (J2000)\\
(MHz) & (MHz) & & & (h) & \\
\hline 
8415   & 50  &  BG0012  & 94-Nov.-17 & 3.0 &   C & 16h28m38.251s
&	+39d33$^{\prime}$04.22$^{\prime\prime}$	\\
8415   & 50  &  GG0038  & 00-Feb.-26 & 3.8 &   BC &16h28m38.243s
&	+39d33$^{\prime}$04.19$^{\prime\prime}$  \\
1665 & 50 & GG0005 & 91-Jun.-18/19 & 8.5 & A & 16h28m38.232s & +39d33$^{\prime}$04.14$^{\prime\prime}$ \\
\hline
\end{tabular}
\end{table*}

\begin{figure*}
\centering
     \begin{tabular}{c c}
     \begin{overpic}[width=0.49\textwidth, trim=50 20 50 20, clip]{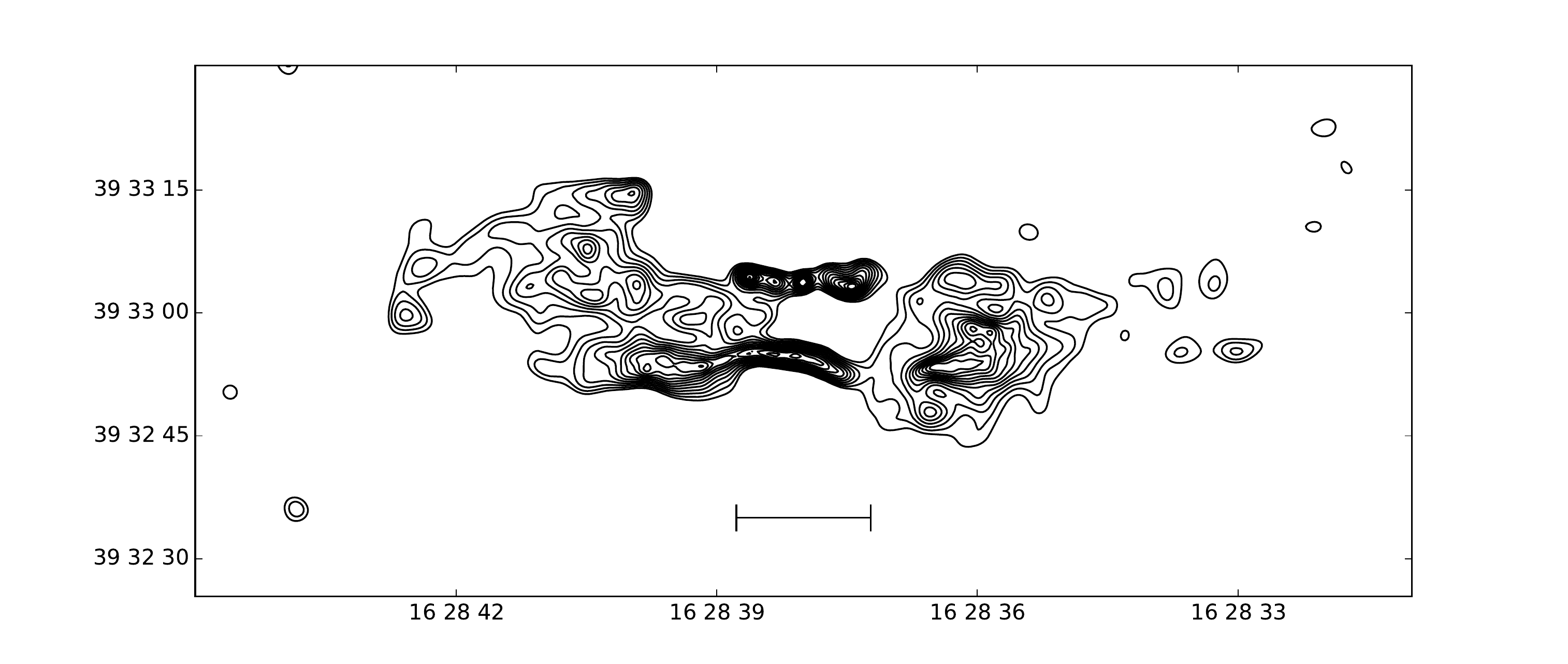}
         \put(47.3,4.5){\tiny 10kpc}
     \end{overpic} &
     \begin{overpic}[width=0.49\textwidth, trim=50 20 50 20, clip]{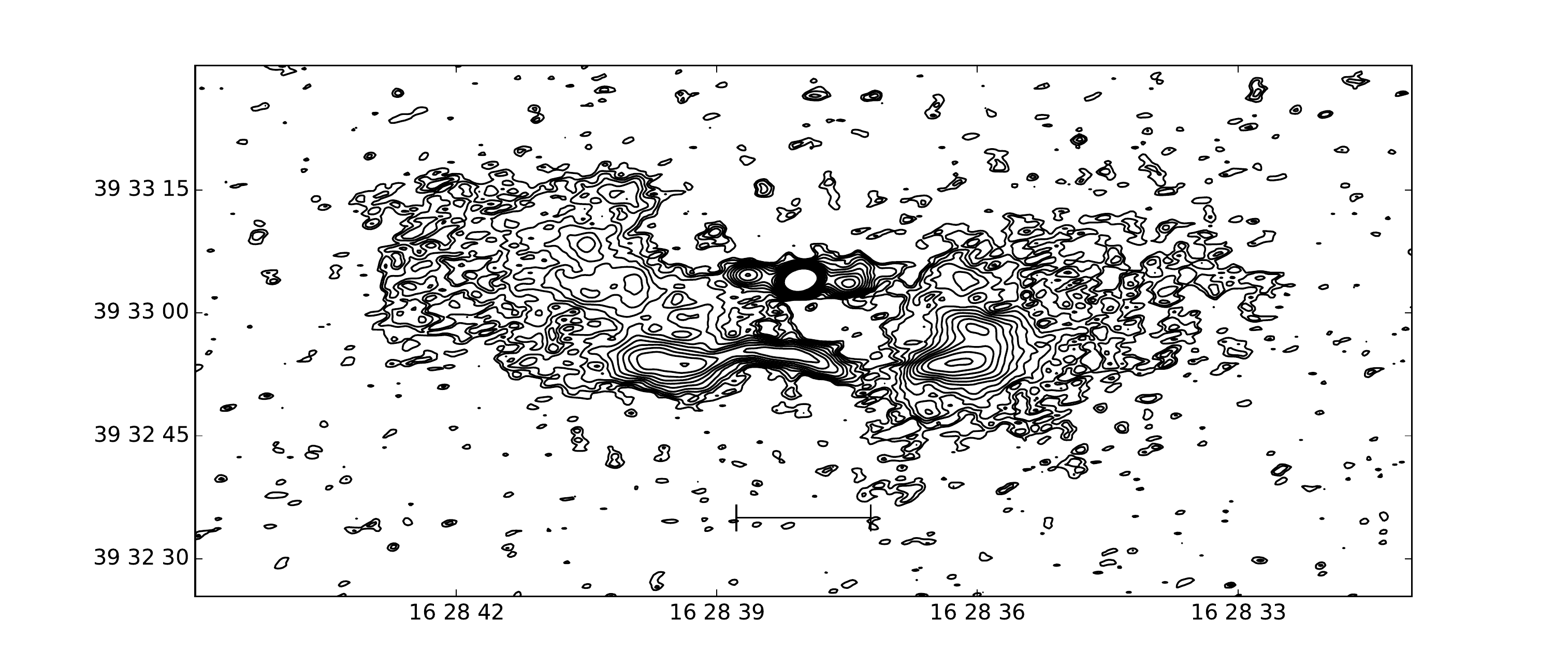}
         \put(47.3,4.5){\tiny 10kpc }
     \end{overpic} \\
     \begin{overpic}[width=0.49\textwidth, trim=50 20 50 20, clip]{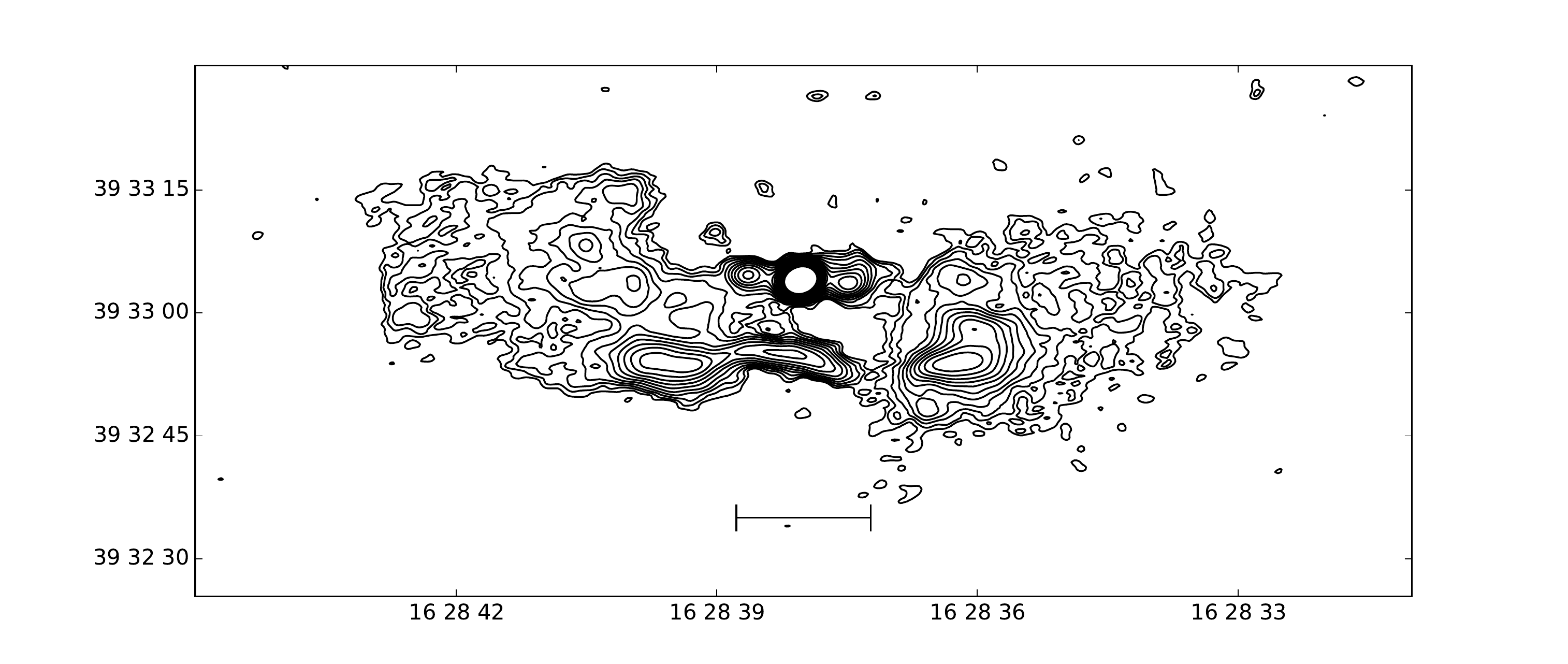}
         \put(47.3,4.5){\tiny 10kpc}
     \end{overpic} &
     \begin{overpic}[width=0.49\textwidth, trim=50 20 50 20, clip]{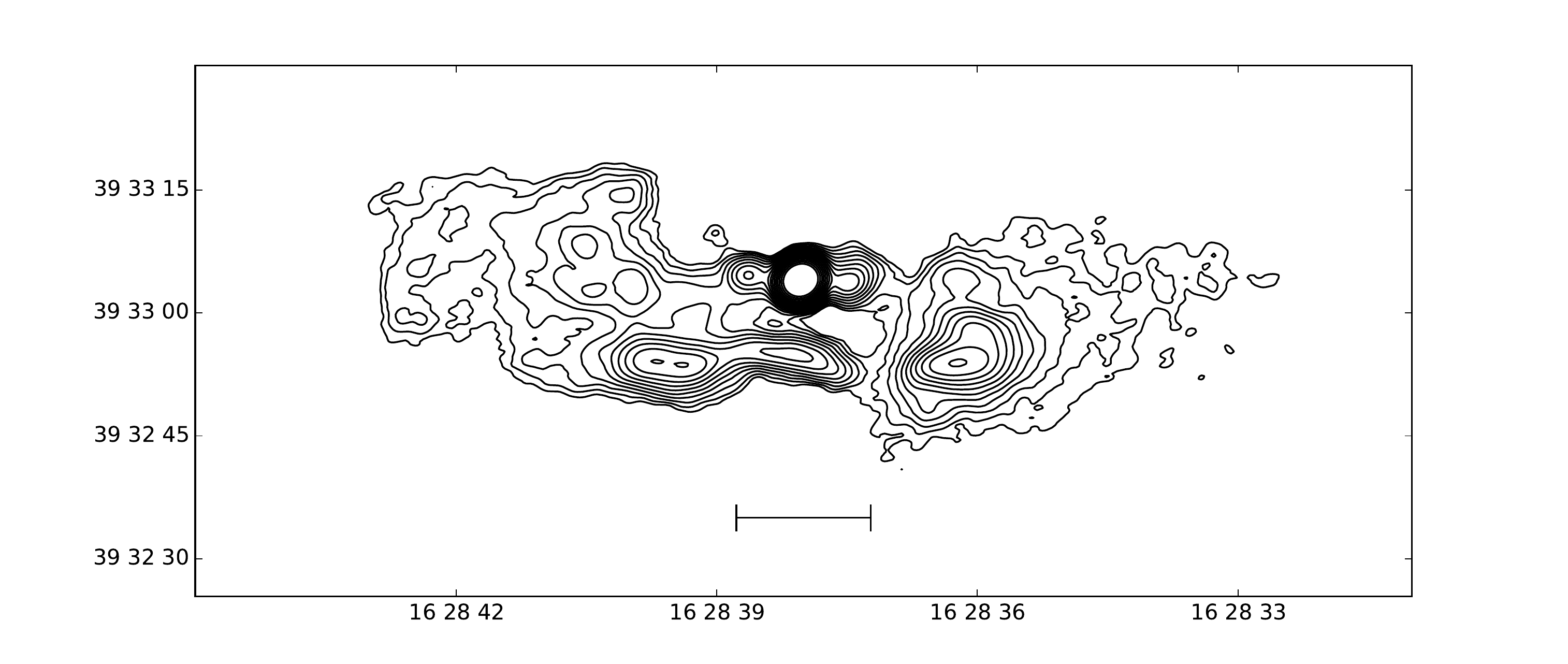}
         \put(47.3,4.5){\tiny 10kpc }
     \end{overpic} \\
     \end{tabular}
 \caption{Total intensity contours of 3C\,388 at 8415 MHz. The top-left panel shows the fast\textsc{Resolve} image, the other panels show the MS-CLEAN images at uniform (top right), intermediate (bottom left) and natural (bottom right) weighting. 
 Contour lines start at $5.2\times 10^5 \mathrm{Jy}/\mathrm{rad}^2$ and increase by factors of $\sqrt{2}$.
 The coordinates are J2000 declination on the y-axis and J2000 right ascension on the x-axis.}
 \label{fig:A2199_contour}
\end{figure*}

\begin{figure*}
\centering
     \begin{tabular}{c c}
     \includegraphics[width=0.49\textwidth, trim=0 40 0 265, clip]{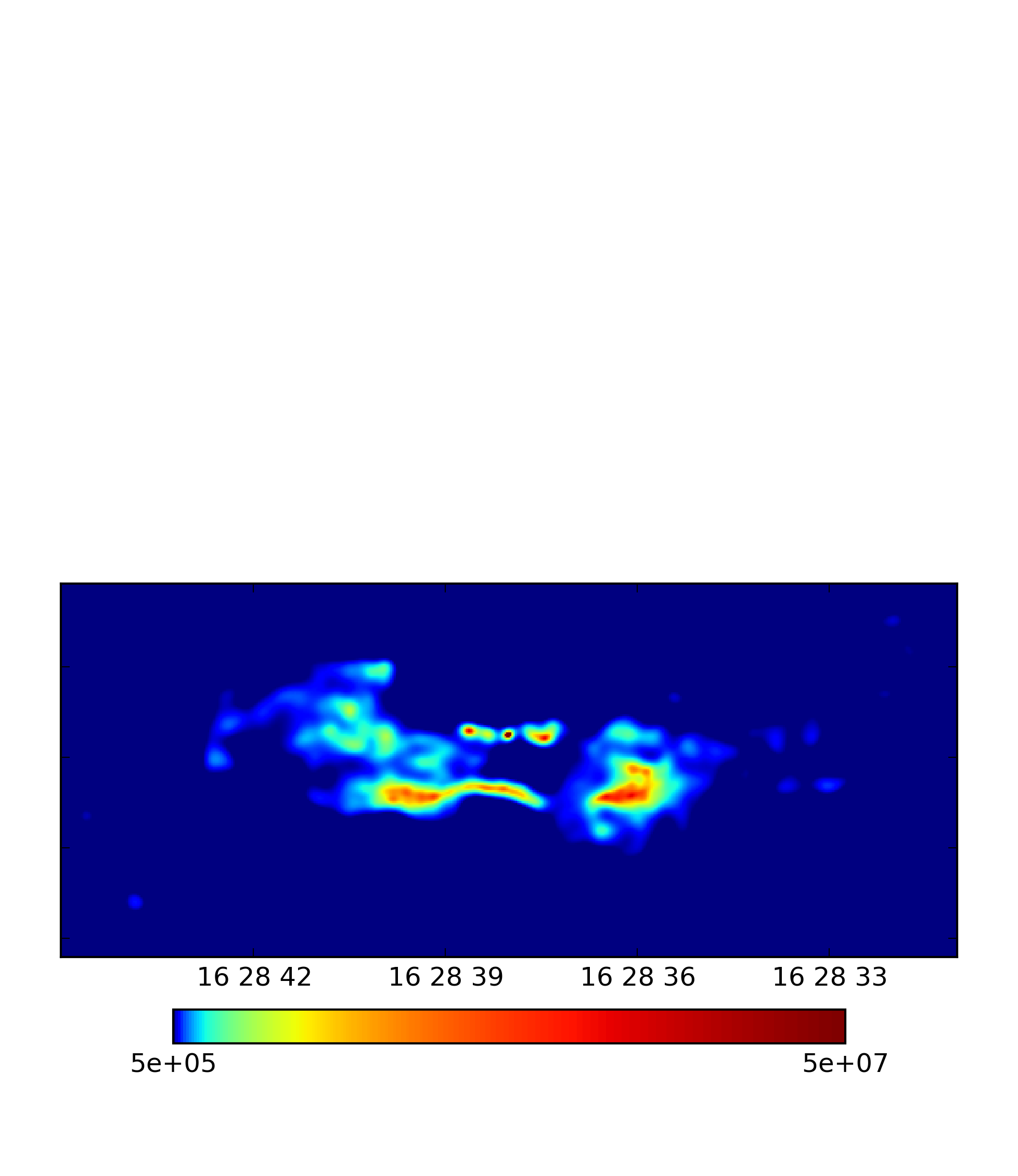} &
     \includegraphics[width=0.49\textwidth, trim=0 40 0 265, clip]{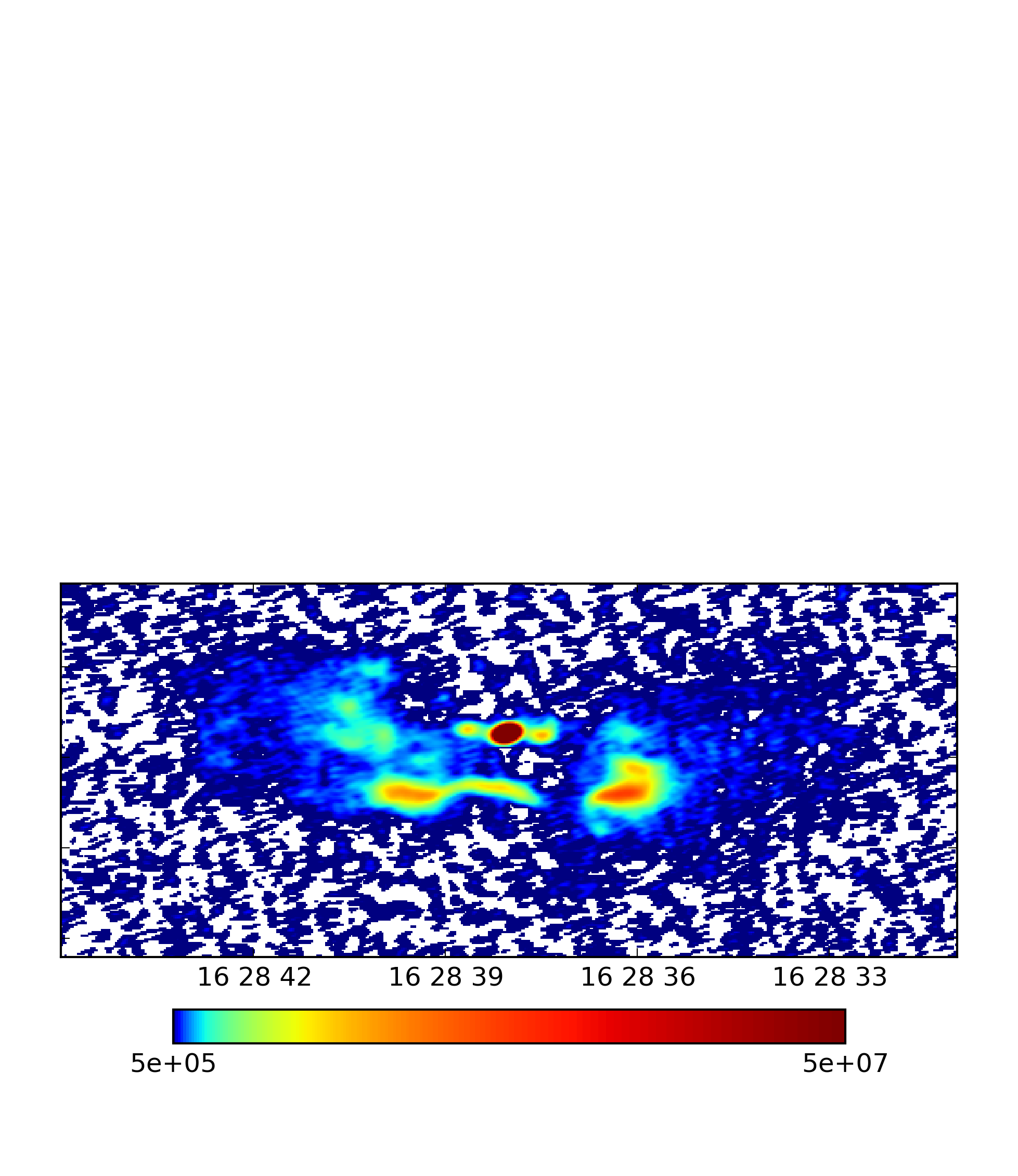} \\
     \includegraphics[width=0.49\textwidth, trim=0 40 0 265, clip]{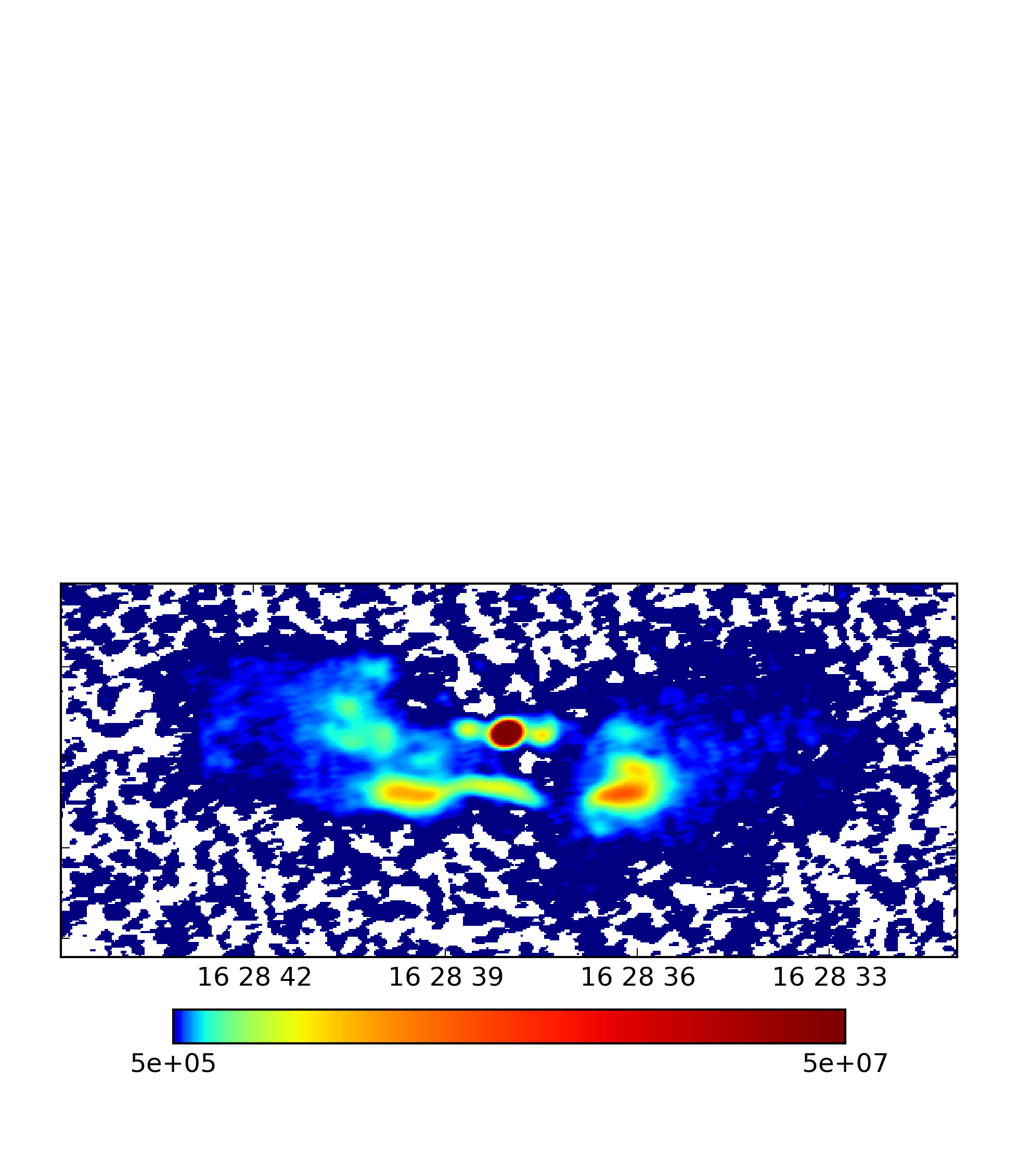} &
     \includegraphics[width=0.49\textwidth, trim=0 40 0 265, clip]{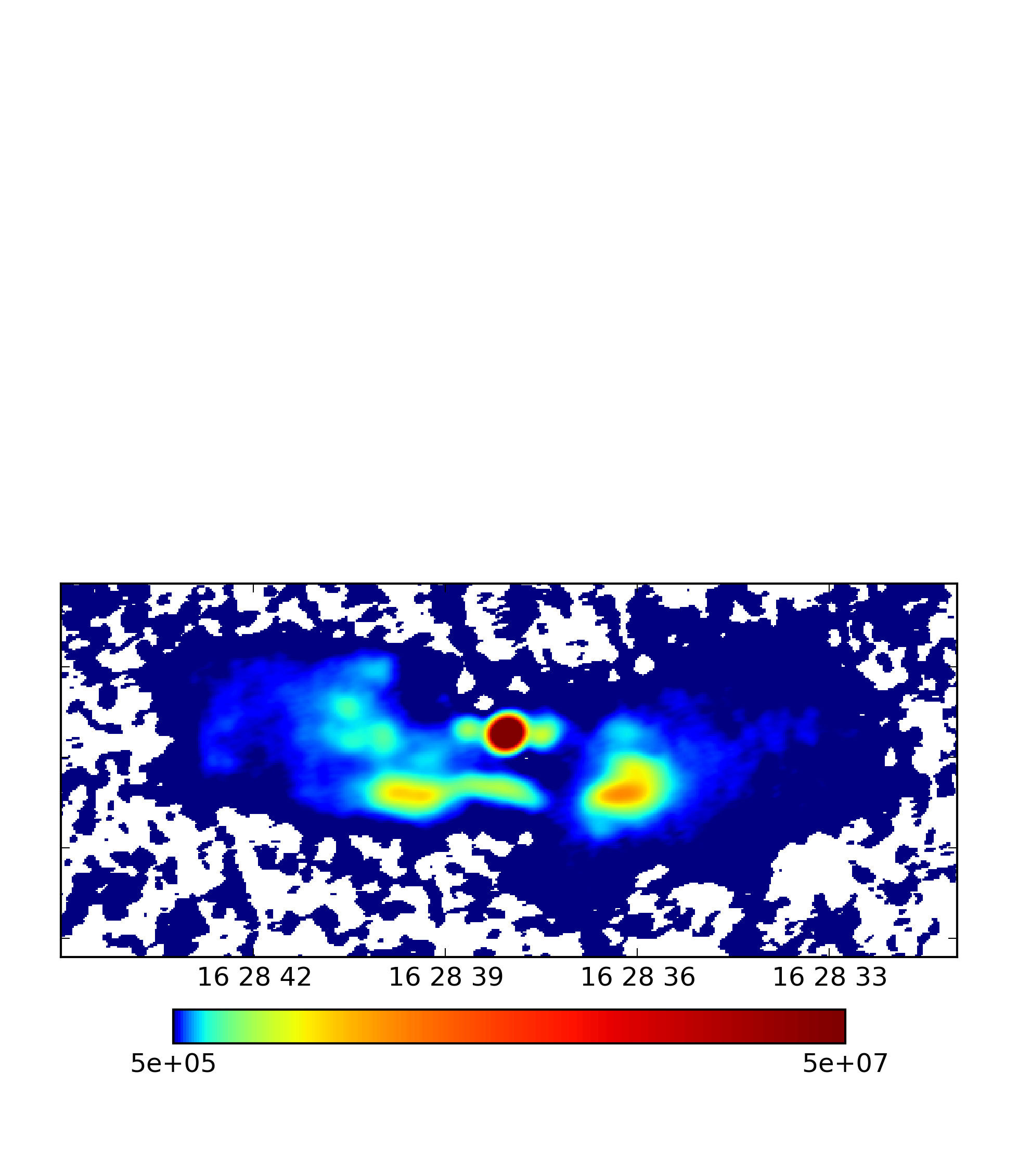} \\
     \end{tabular}
 \caption{Total intensity of 3C\,388 at 8415 MHz. The top-left panel shows the fast\textsc{Resolve} image, the other panels show the MS-CLEAN images at uniform (top right), intermediate (bottom left) and natural (bottom right) weighting. The units are $\mathrm{Jy}/\mathrm{rad}^2$. The white regions in the MS-CLEAN image are regions with negative intensity. The panels show the same field of view as in Fig.~\ref{fig:A2199_contour}}
 \label{fig:A2199_plot}
\end{figure*}

\begin{figure*}
\centering
     \begin{tabular}{c}
     \includegraphics[width=0.98\textwidth, trim=0 97 0 265, clip]{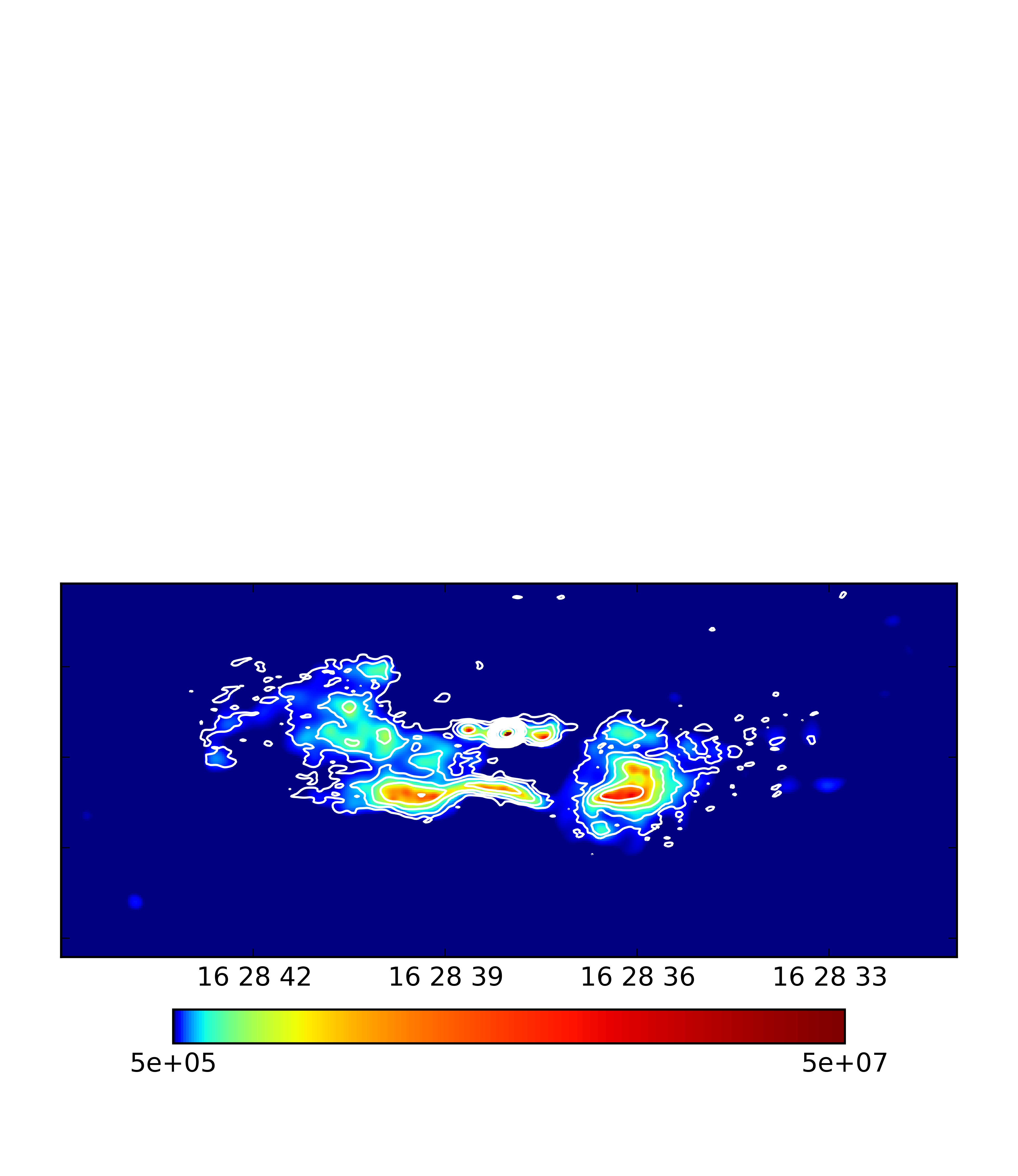} 
     \end{tabular}
 \caption{Superposition of the fast\textsc{Resolve} and MS-CLEAN image of 3C\,388 at 8415 MHz. The color plot shows the fast\textsc{Resolve} image, the contour lines show the MS-CLEAN image (uniform weighting). The color code, and the field of view are the same as in Fig.~\ref{fig:A2199_plot}. The contour lines start at $10.4\times 10^5 \mathrm{Jy}/\mathrm{rad}^2$ and increase by factors of 2.}
 \label{fig:A2199_overplot}
\end{figure*}

In Figs.~\ref{fig:A2199_contour} and \ref{fig:A2199_plot} we compare the fast\textsc{Resolve} reconstruction with the MS-CLEAN reconstructions using different weightings. Additionally we show a superposition of the fast\textsc{Resolve} image with the uniform weighting MS-CLEAN image in Fig.~\ref{fig:A2199_overplot}.
The reconstructions agree in their large-scale features. Small-scale features, which are comparable to the size of the beam, show the advantage of fast\textsc{Resolve}. The brightest structures are sharper and more detailed than in the MS-CLEAN image with uniform weighting, while the faint structures are almost as smooth as in the natural weighting image.
In fast\textsc{Resolve} the effective resolution is chosen automatically through the interplay of prior and likelihood, i.e. structures which are noise dominated are washed out. 
This allows fast\textsc{Resolve} to adapt the resolution locally, depending on the strength of the data for that specific location \citep[see the detailed discussion in][]{Junklewitz-2016}. Therefore, the faint regions are smoother in fast\textsc{Resolve} whereas the bright regions are sharper. One can find many features which are hinted at in the MS-CLEAN image to be much more detailed in the fast\textsc{Resolve} image.
This effect of superior resolution compared to (MS) CLEAN will be thoroughly discussed in Junklewitz~et~al.~(in prep.) using VLA data of the radio galaxy Cygnus A.
Furthermore, the MS-CLEAN image exhibits negative regions, whereas the fast\textsc{Resolve} image is strictly positive. 

\begin{table}
\caption{Flux of 3C\,338 derived with the different approaches.}
\label{tab:fluxes}
\centering
\begin{tabular}{c c c}
\hline\hline
weighting/mask & MS-CLEAN flux & fast\textsc{Resolve} flux\\
\hline
uniform & $174\,\mathrm{mJy}$ &  $162^{+20}_{-15}\,\mathrm{mJy}\vphantom{\Big(}$ \\
intermediate & $174\,\mathrm{mJy}$ & $166^{+20}_{-17}\,\mathrm{mJy}\vphantom{\Big(}$ \\
natural & $172\,\mathrm{mJy}$ & $166^{+21}_{-16}\,\mathrm{mJy}\vphantom{\Big(}$\\
no mask & -- & $187^{+35}_{-24}\,\mathrm{mJy}\vphantom{\Big(}$ \\
\hline
\end{tabular}
\end{table}

In Table \ref{tab:fluxes} we list the total flux estimates obtained with fast\textsc{Resolve} and MS-CLEAN. The estimates are derived by masking out all the regions with a flux below three times the noise of the MS-CLEAN images and integrating\footnote{All of the integration in this paragraph are performed over a square with an edge length of $2'33.6''$ centered on the pointing and including all the sources of emission. This is to avoid artifacts towards the edge of the primary beam which can be present in both MS-CLEAN and fast\textsc{Resolve}.} the remaining flux for each MS-CLEAN image. \corr{The noise levels of the MS-CLEAN images were $0.02\mathrm{mJy}/\mathrm{beam}$ for uniform and $0.015\mathrm{mJy}/\mathrm{beam}$ for intermediate and natural weighting.} This yields three different masks for which we list the corresponding fluxes of the fast\textsc{Resolve} image, as well. Since the fast\textsc{Resolve} image has no negative regions there is no imperative need for a mask. We therefore list the total flux of the fast\textsc{Resolve} image, too.
All derived MS-CLEAN fluxes agree with the corresponding fast\textsc{Resolve} fluxes within the credibility intervals. Since fast\textsc{Resolve} separates point-like and diffuse flux, we can state an estimate for the total point-like flux. It is $84\,\mathrm{mJy}$. An uncertainty estimate for the point-like flux is not provided by fast\textsc{Resolve}.

\begin{figure*}
\centering
     \begin{tabular}{c c}
     \includegraphics[width=0.49\textwidth, trim=0 40 0 268, clip]{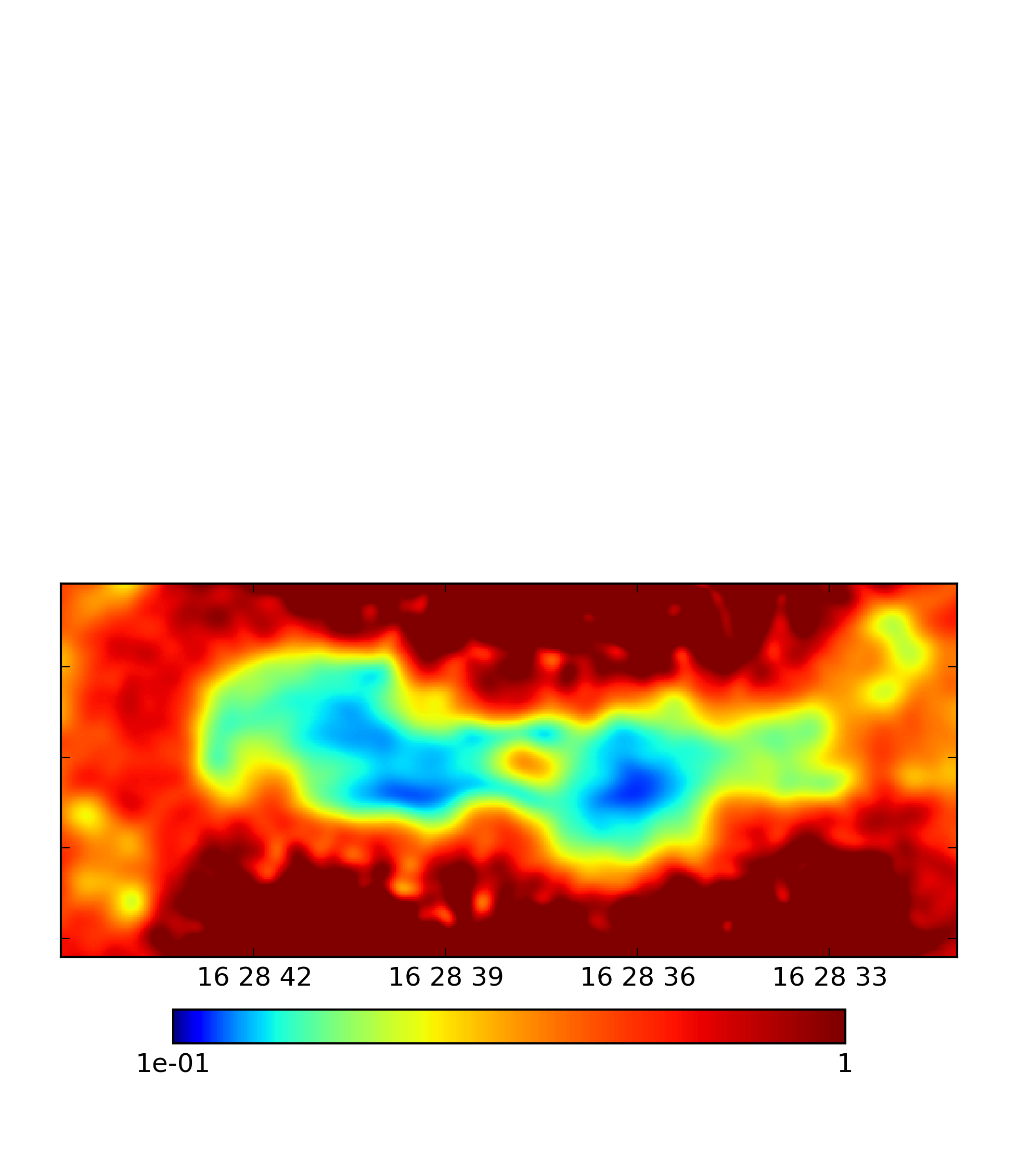} &
     \includegraphics[width=0.49\textwidth, trim=0 40 0 268, clip]{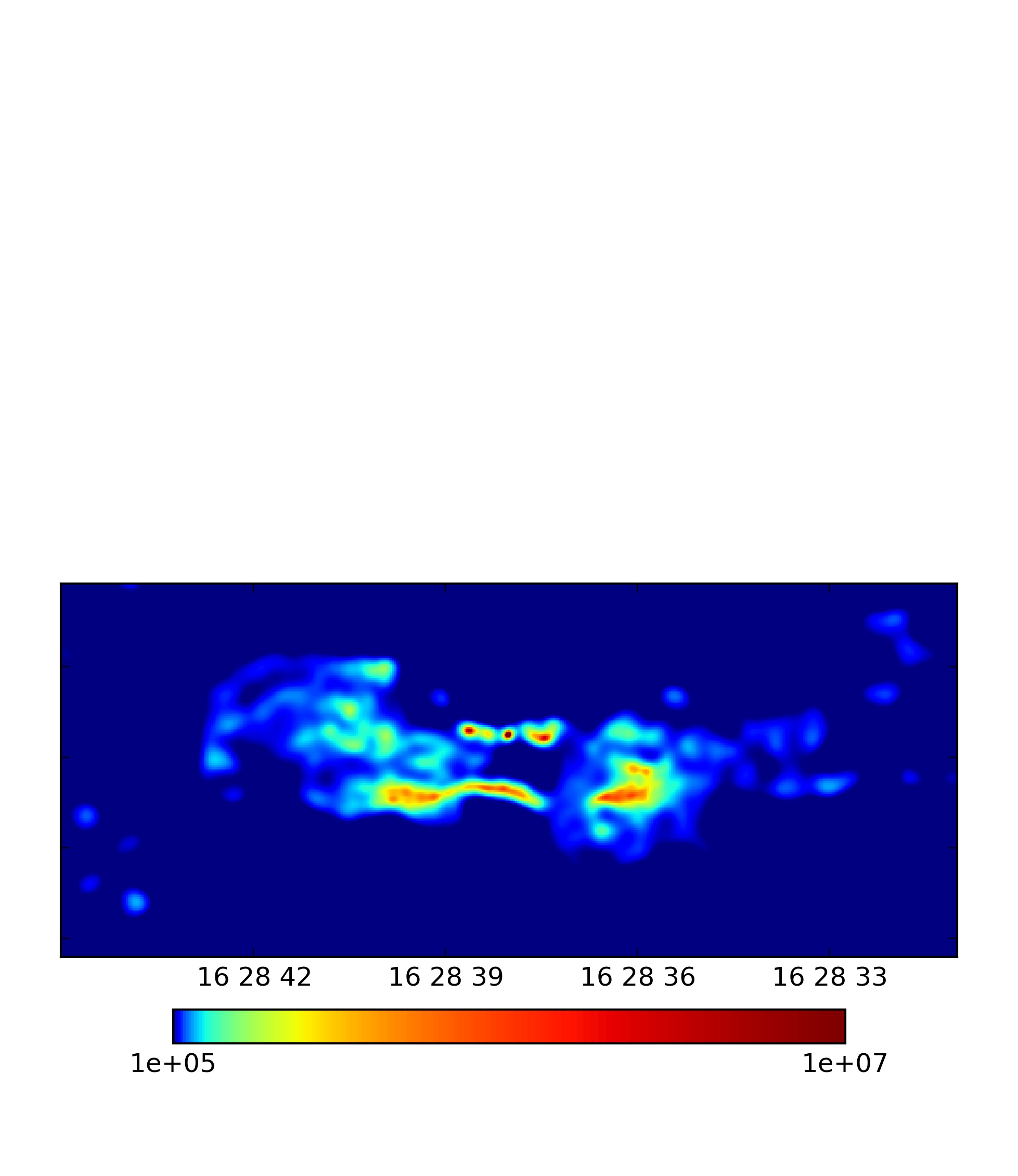} \\
     \end{tabular}
 \caption{Relative (left panel) and absolute (right panel) uncertainty of the fast\textsc{Resolve} image of 3C\,388 at 8415 MHz. The same field of view as in Fig.~\ref{fig:A2199_contour} is shown. \corr{Units of the absolute uncertainty are $\mathrm{Jy}/\mathrm{rad}^2$.}}
 \label{fig:A2199_uncertainty}
\end{figure*}

Fig.~\ref{fig:A2199_uncertainty} shows the estimate of the relative \corr{and absolute} uncertainty ($1\sigma$ credibility interval) of the fast\textsc{Resolve} image, a quantity that is not accessible for a MS-CLEAN reconstruction. We see that the relative uncertainty is larger for faint regions and smaller for bright regions. This is because bright regions have a larger impact on the data and are therefore easier to separate from the noise. \corr{The absolute uncertainty follows the reconstructed intensity strongly in morphology, but with a lower dynamic range. In summary, regions with high flux have lower relative, but higher absolute uncertainty than regions with low flux.}
It is important to note that this is only an estimate of the credibility interval using a saddle point approximation not the result of a full sampling of the posterior distribution. Thus it is likely that the true credibility interval is larger. fast\textsc{Resolve} provides an estimate of the likelihood variances as well. The variance\footnote{Note that $\sigma^2$ is an estimate of the statistical uncertainty of the measured visibilities, not of the reconstructed flux.} of the visibilities is uniformly ($t=0$ in Eq.~\eqref{eq:sigma-regu}) estimated as $\sigma^2 = \left( 13.6\,\mathrm{mJy} \right)^2$.
Thus the average signal-to-noise ratio is $S\!N\!R\approx37$ for the point-like emission, $S\!N\!R\approx1$ for the diffuse emission and $S\!N\!R\approx40$ for the combined emission. The point source is much stronger imprinted in the data than the diffuse flux, since a sharp peak excites all Fourier modes (see Appendix \ref{sec:SNR_formula} for the $S\!N\!R$ definition used in this work).

\subsubsection{Computing time}
\label{sec:run-time}

Comparing the computing time of fast\textsc{Resolve} to MS-CLEAN and \textsc{Resolve} is not straight forward.
\textsc{Resolve} and fast\textsc{Resolve} do not require user input during the runtime.
MS-CLEAN typically needs repeated user input (interactive CLEANing) to select the appropriate regions in which flux is expected. Therefore, its performance cannot be characterized in computation time alone. 
In fast\textsc{Resolve} and \textsc{Resolve} the convergence criteria of the minimization in Eq.~\eqref{eq:MAP-equations-m} play a decisive role for the run-time and since they use different minimization schemes it is not entirely clear how to compare these criteria. \textsc{Resolve} uses a gradient descent with Wolfe conditions while fast\textsc{Resolve} uses the \mbox{L-BFGS-B} algorithm \citep[][]{LBFGS-1995}. However, in Appendix~\ref{sec:mock_diffuse} we find the computation time of fast\textsc{Resolve} to be roughly 100 times shorter compared to \textsc{Resolve} in our mock test.
What we compare here is the time it takes to evaluate the likelihood and its gradient. Any efficient minimization algorithm will have to evaluate both repeatedly. After removing flagged data points the data set used in the previous section consists of 750489 $uv$-points. Evaluating the likelihood in \textsc{Resolve} took 11.9 seconds using a single core of an Intel Xeon E5-2650 v3 @ 2.30GHz. Using the same setup the likelihood in fast\textsc{Resolve} took 52 milliseconds to evaluate. The gradients took 27.1 seconds in \textsc{Resolve} and 77 milliseconds in fast\textsc{Resolve}. Combined with the evaluation of the prior, which is the same in \textsc{Resolve} and fast\textsc{Resolve}, this yields speed up by a factor of over 100. The fast\textsc{Resolve} likelihood scales as $\mathcal{O}(N_\mathrm{pix} \log N_\mathrm{pix})$ and the \textsc{Resolve} likelihood scales as $\mathcal{O}(N_\mathrm{pix} \log N_\mathrm{pix} + WN_\mathrm{uv})$, where $W$ is a precision parameter and $N_\mathrm{uv}$ is the amount of $uv$-points. The prior scales as $\mathcal{O}(N_\mathrm{pix} \log N_\mathrm{pix})$ for both algorithms. Therefore, the speedup factor should increase with the amount of $uv$-points and descrease with the amount of image pixels.

In the fast\textsc{Resolve} run presented in the previous chapter it took 3 hours and 22 minutes on one CPU core to calculate the map $m$ and 4 hours and 8 minutes on 2 cores to calculate the uncertainty map.
Applying MS-CLEAN to the data took approximately $30$ minutes of interactive work using one core of an AMD FX-8350 @ 4GHz.
This is still a big time difference -- it is roughly a factor of 7 for the image alone -- but it makes it feasible to run fast\textsc{Resolve} on a conventional notebook. The calculation of the fast\textsc{Resolve} image was performed using rather strict convergence criteria to be on the safe side. By relaxing the convergence criteria we managed to decrease the computation time of the map to 50 minutes without significantly changing the result.
Another factor is the FFT implementation. Currently, fast\textsc{Resolve} uses the numpy FFT implementation. By using the FFTW \citep[][]{FFTW05}, there could be another speedup by a factor of 2 and even more if one uses more than one core (which is not possible in the numpy FFT).

\subsubsection{A2199 -- 1665MHz}

We tested fast\textsc{Resolve} on several other data sets of 3C\,388 at frequencies between 1.6 GHz and 8.4 GHz and compared the results to the corresponding CLEAN images. Most comparisons lead us to the same conclusions as the 8415 MHz data in the previous section. In this section we show the results of the data set for which fast\textsc{Resolve} performed worst in our opinion. The data were recorded at 1665 MHz, but the antenna configuration was different than in the previous section so that similar scales are probed by both, the 1665 MHz and the 8415 MHz data. The details of the data sets can be found in Tab.~\ref{tab:observations}.

\begin{figure*}
\centering
     \begin{tabular}{c c}
     \includegraphics[width=0.49\textwidth, trim=0 40 0 265, clip]{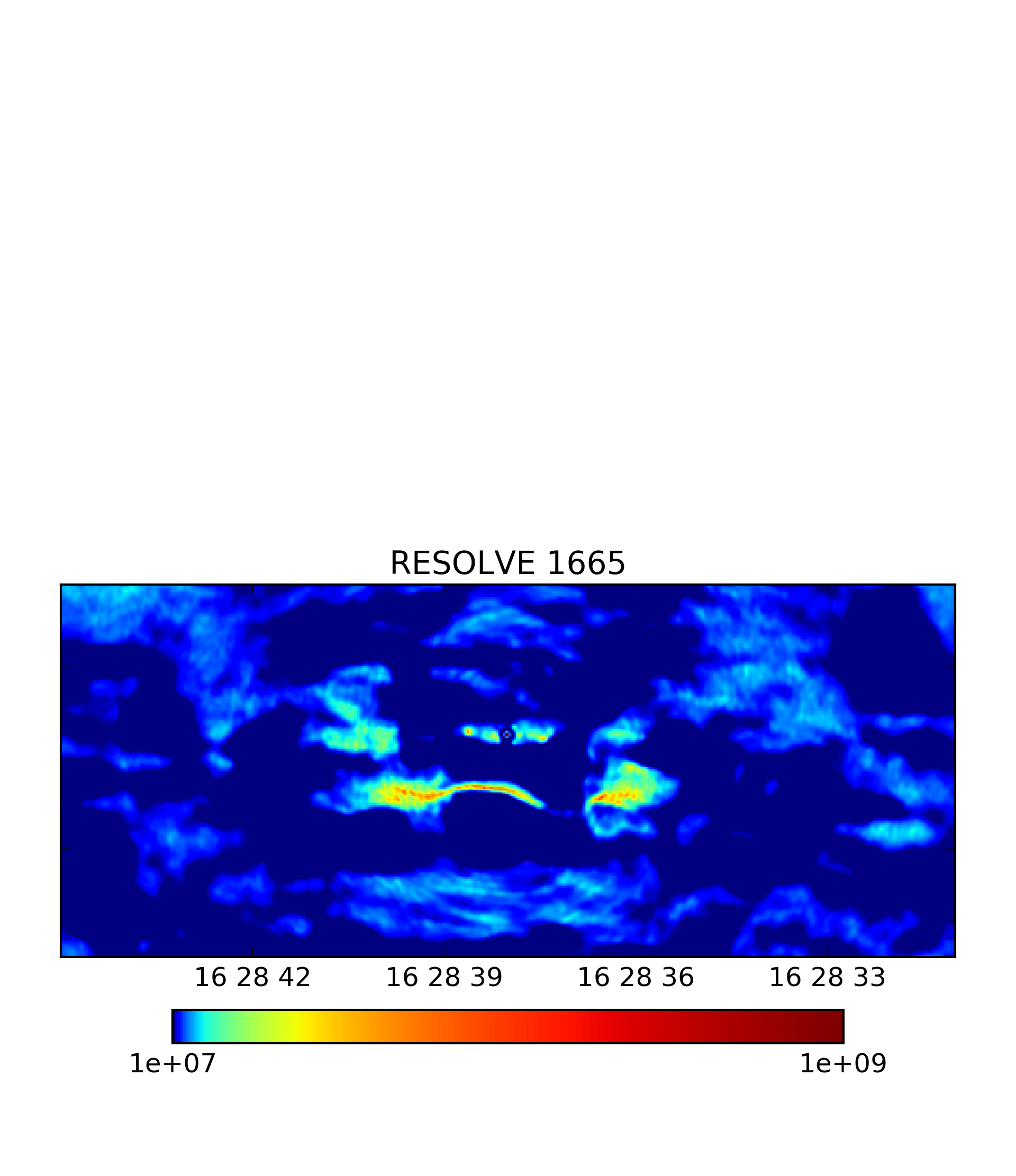} &
     \includegraphics[width=0.49\textwidth, trim=0 40 0 265, clip]{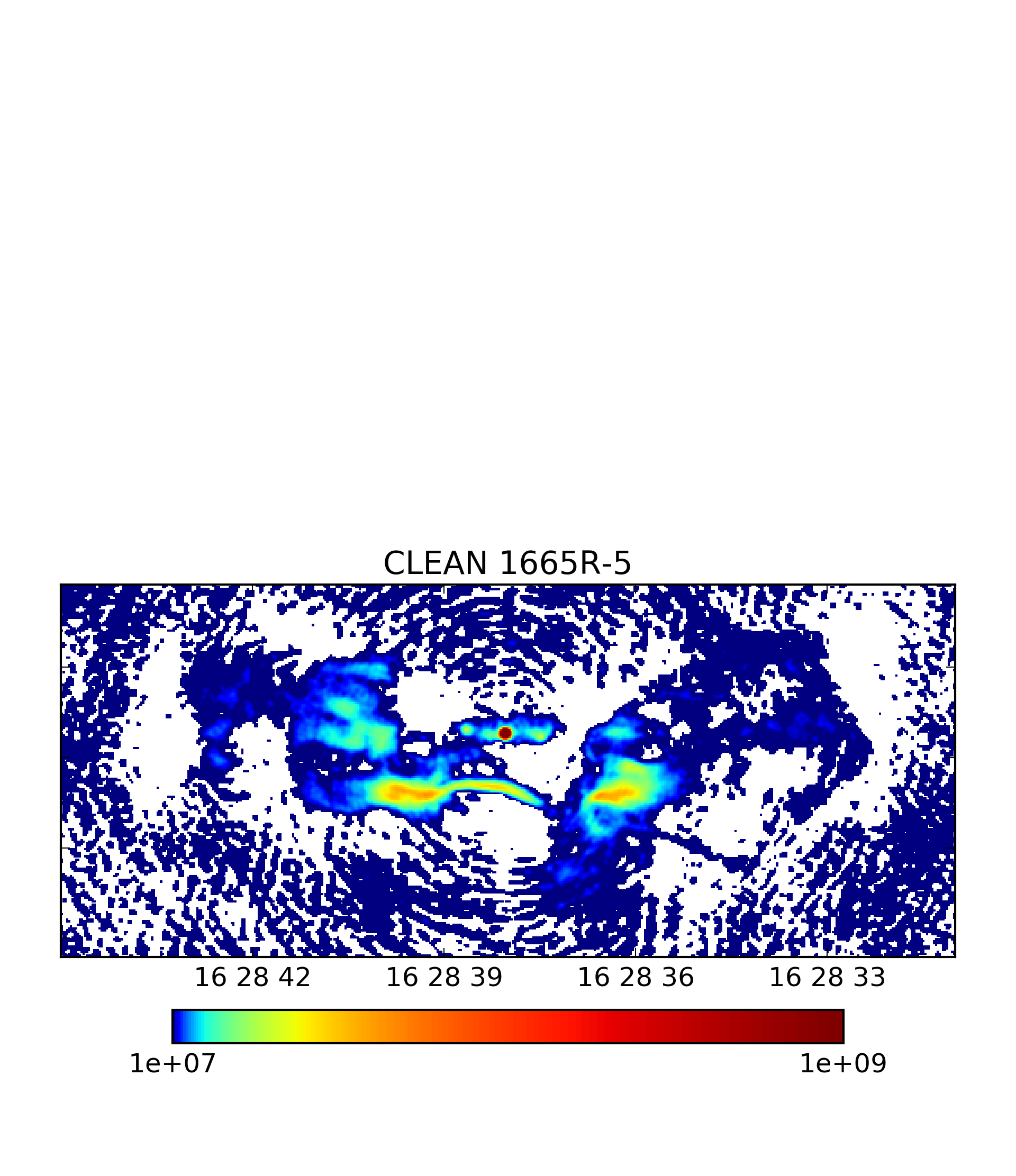} \\
     \end{tabular}
 \caption{Total intensity of 3C\,388 at 1665 MHz. The left panel shows the fast\textsc{Resolve} image, the right panel shows the CLEAN image at uniform weighting. The units are $\mathrm{Jy}/\mathrm{rad}^2$. The white regions in the CLEAN image are regions with negative intensity. The panels show the same field of view as in Fig.~\ref{fig:A2199_contour}}
 \label{fig:A2199_1665_plot}
\end{figure*}

In Fig.~\ref{fig:A2199_1665_plot} we show the fast\textsc{Resolve} image and the CLEAN\footnote{\corr{Here we used standard CLEAN, not MS-CLEAN.}} image at uniform weighting. One can see that the fast\textsc{Resolve} image shows significant flux in regions where there should be no emission and that there is some artificial clumpiness on small scales. Furthermore, there is a ring of low emission around the nucleus, which is probably an artifact. Such a ring might be the result of overfitting the point-like nucleus, but we could not eliminate this artifact even by manually adjusting the size and strength of the point-like contribution to the image. The ring seems to be supported by the data which leads us to the suspicion that it is due to an imperfect calibration. Such a miscalibration could also explain the other artifacts in the fast\textsc{Resolve} image. Generally, it is fair to say that fast\textsc{Resolve} is more sensitive to calibration problems than CLEAN. However, even with the adressed problems the fast\textsc{Resolve} image still compares nicely to the CLEAN image in our opinion. They both resolve structures to a similar degree and they both exhibit a number of artifacts (eg.~negative flux regions in the CLEAN image).

\section{Discussion \& Conclusion}
\label{sec:summary}

We presented fast\textsc{Resolve}, a Bayesian imaging algorithm for aperture synthesis data recorded with a radio interferometer. It is much cheaper than \textsc{Resolve} in computation time without losing the key advantages \textsc{Resolve} has compared to CLEAN (optimality for diffuse emission and uncertainty propagation).
Furthermore, fast\textsc{Resolve} is capable of reconstructing point sources and estimating the measurement variance of the visibilities. The estimation of the measurement variances is completely a new approach in aperture synthesis. It permits the application of Bayesian imaging schemes which rely on an accurate description of the measurement variances, which is often not provided for radio interferometric data.

We tested fast\textsc{Resolve} on archival VLA data of Abell 2199. We showed reconstructions at $1665\,\mathrm{MHz}$ and $8415\,\mathrm{MHz}$. For the $8415\,\mathrm{MHz}$ data fast\textsc{Resolve} could resolve more detailed structures than CLEAN in uniform weighting while introducing fewer artifacts than CLEAN in natural weighting (especially no regions negative brightness). This behavior could be observed at most of the other frequencies we used for testing.
We showed the $1665\,\mathrm{MHz}$ data set as a negative example. For this data set fast\textsc{Resolve} slightly overfitted small-scale structures and introduced artifacts that put additional flux into empty regions. However, the overall quality of the image was still comparable to CLEAN at uniform weighting. Therefore, fast\textsc{Resolve} could be regarded as a replacement or at least supplement to CLEAN.

The speed-up of fast\textsc{Resolve} is achieved by introducing an approximation scheme in which the data are gridded onto a regular Fourier grid to allow for a quicker evaluation of the likelihood and its derivative.
This approximation scheme could also serve as a general compression scheme in aperture synthesis, if the amount of $uv$ points exceeds the amount of grid points of the desired image.
This will be easily the case for measurements by the upcoming Square Kilometre Array (SKA). In such cases, one could save $j$ and the diagonal description of $M$ (see Sec.~\ref{sec:Mdiag}). Both need the same amount of space as the final image. Calibration, flagging, as well as the pixelization cannot be easily changed afterwards (if at all), but one preserves a description of the likelihood and can still apply different visibility dependent imaging and modeling schemes. \corr{However, in order to be applicable for SKA data a generalization of the fast\textsc{Resolve} approximation, that does not depend of the flat sky approximation, has to be developed.}

fast\textsc{Resolve} will be included into the main \textsc{Resolve} package as an optional feature.
A hybrid version in which the result of fast\textsc{Resolve} is used to speed up a final \textsc{Resolve} run will also be implemented. Within this inclusion the multi-frequency capability of \textsc{Resolve} will also be enabled for fast\text{Resolve}.
The open source Python code is publicly available\footnote{
\begin{tabular}{r l}
ASCL: &  \url{http://ascl.net/1505.028}\\
github: & \url{http://github.com/henrikju/resolve}
\end{tabular}
}.
As has been mentioned earlier, there is another add-on under development which will enable a more consistent treatment of point-sources.
All of these efforts will be joined to make the \textsc{Resolve} algorithm optimal for both, diffuse and point sources while keeping the computation time minimal.

\begin{acknowledgements}
 We wish to thank Martin Reinecke for his help with the C implementation of the gridding function. The calculations were realized using the \textsc{NIFTy}\footnote{\url{http://www.mpa-garching.mpg.de/ift/nifty/}} package by \cite{Selig-2013}. Some of the minimizations in this work were performed using the \mbox{L-BFGS-B} algorithm \citep[][]{LBFGS-1995}. This research has been partly supported by the DFG Research Unit 1254 and has made use of NASA’s Astrophysics Data System.
\end{acknowledgements}

\begin{appendix}
 
 \section{Aliasing}
 \label{sec:aliasing}

  The visibilities are calculated by a Fourier transform of the product between intensity and primary beam Eq.~\eqref{eq:interferometer}. In practice we are working on a pixelized grid where the integrals become sums,
  \begin{equation}
  \begin{split}
   V(u,v) & = \sum\limits_{x \in X} \Delta x \,e^{-2\pi i\, u x} \sum\limits_{y \in Y} \Delta y \,e^{-2\pi i\, v y}\ B(x,y) \, I(x,y),
  \end{split}
  \end{equation}
 with 
 \begin{equation}
  \begin{split}
   X & = \left\{ j \Delta x\ | \ -N_x/2\leq j < N_x/2 \right\},\\ 
   Y & = \left\{ j \Delta y\ | \ -N_y/2\leq j < N_y/2 \right\}.
  \end{split}
 \end{equation}
 Inserting a discrete Fourier transformation and inverse transformation before $B(x,y) \, I(x,y)$ yields
 \begin{equation}
  \begin{split}
   V(u,v)  = & \sum\limits_{x \in X} \Delta x \,e^{-2\pi i\, u x} \sum\limits_{k \in K} \Delta k \,e^{2\pi i\, k x} \\
             & \sum\limits_{y \in Y} \Delta y \,e^{-2\pi i\, v y} \sum\limits_{q \in Q} \Delta q \,e^{2\pi i\, q y}  \ \tilde{I}(k,q)
  \end{split}
 \end{equation}
with
 \begin{equation}
  \tilde{I}(k,q) = \sum\limits_{x \in X}\Delta x\sum\limits_{y \in Y}\Delta y\, e^{-2\pi i (kx + qy)}\, B(x,y) \, I(x,y)
 \end{equation}
and
 \begin{equation}
  \begin{split}
   K  = \left\{ j \Delta k\ | \ -N_x/2\leq j < N_x/2 \right\},\quad & \Delta k = \frac{1}{N_x\Delta x},\\ 
   Q  = \left\{ j \Delta q\ | \ -N_y/2\leq j < N_y/2 \right\},\quad & \Delta q = \frac{1}{N_y\Delta y}.
  \end{split}
 \end{equation}
 Summing over $x$ and $y$ first makes this
 \begin{equation}
  \begin{split}
   V(u,v) & = \sum\limits_{k \in K} \Delta k \sum\limits_{q \in Q} \Delta q\ G(u,v;k,q)\, \tilde{I}(k,q),
  \end{split}
 \end{equation}
 with the gridding operator
 \begin{equation}
  \begin{split}
   G(u,v;k,q) =\ & \Delta x\,\frac{2 i \, \sin\!\left( \pi (k - u ) /\Delta k \right)} {\mathrm{e}^{2 i \, \pi(k-u)/(N_x \Delta k)}-1} \times \\
             & \Delta y\, \frac{2 i \, \sin\!\left( \pi (q - v ) /\Delta q \right)}{\mathrm{e}^{2 i \, \pi(q-v)/(N_y \Delta q)}-1}.
  \end{split}
  \label{eq:aliasing_exact}
 \end{equation}
To emphasize the difference between the unstructured set of $uv$-points and the structured set of $k$ and $q$ points, we shorten the notation of $G$ to
\begin{equation}
 G_j(k,q) \equiv G(u_j,v_j;k,q).
\end{equation}
If $N_x$ and $N_y$ are large, $G$ can be approximated as
\begin{equation}
 \begin{split}
   G_j(k,q) \approx & \frac{1}{\Delta k\, \Delta q} \mathrm{sinc}\!\left( \frac{k-u_j}{\Delta k} \right) \mathrm{sinc}\!\left( \frac{q-v_j}{\Delta q} \right),
 \end{split}
 \label{eq:aliasing_approx}
\end{equation}
with $\mathrm{sinc}(x) := \sin(\pi x)/(\pi x)$.
To project the visibilities $V_j \equiv V(u_j,v_j)$ into the regular space the adjoint of $G$ has to be applied:
\begin{equation}
 V(k,q) = \sum\limits_{j} G^*_j(k,q)\, V_j.
\end{equation}
The inverse (fast) Fourier transform of $V(k,q)$ yields the same result as the sum of the inverse (direct) Fourier transform of $V_j$.

\corr{In Eq.~\eqref{eq:aliasing_approx} it is easy to see, that $G_j(k,q)$ falls off with the inverse distance from the maximal pixel in each direction. This behavior is independent of the pixelization as the pixel edge lengths are divided out in the $\mathrm{sinc}$ terms. This allows for a pixelization independent cut-off of the sinc-kernel. It should be noted that state-of-the-art gridding codes do not use a truncated $\mathrm{sinc}$ function to grid the data, but more elaborate schemes, which have a better cut-off behavior, e.g.~the Kaiser-Bessel gridding kernel \citep[][]{Beatty-2005}. For our approximation we achieve better results with a truncated version of \eqref{eq:aliasing_exact}, where we truncate after 5 pixels in each dimension.}
 
 \section{Kullback-Leibler Divergence exremization}
 \label{sec:DKL_minimization}

 In this section we derive the information theoretically optimal approximation of a Gaussian likelihood with linear measurement by a diagonal measurement precision operator.
 The likelihood we want to approximate is of the form
 \begin{equation}
  \mathcal{P}(d|s) = \frac{1}{\left| 2 \pi N \right|^{\frac{1}{2}}} \exp\!\left( -\frac{1}{2} \left(d - Rs\right)^\dagger N^{-1} \left(d - Rs\right) \right).
 \end{equation}
 Combined with a Gaussian prior it yields the posterior
 \begin{equation}
 \begin{split}
 \mathcal{P}(s|d)  = &\left| 2 \pi \left(S^{-1} + R^\dagger N^{-1} R   \right)^{-1}\right|^{-\frac{1}{2}} \\
   \times &\ \mathrm{e}^{ -\frac{1}{2} \left(s - R^\dagger N^{-1} d\right)^\dagger \left( S^{-1} + R^\dagger N^{-1} R  \right) \left(s - R^\dagger N^{-1} d\right)}.
 \end{split}
 \end{equation}
The likelihood enters the posterior via two quantities: the information source $j=R^\dagger N^{-1} d$ and the measurement precision $M=R^\dagger N^{-1} R$. Therefore, a Gaussian likelihood with a linear response always yields a posterior of the form
\begin{equation}
 \begin{split}
  \mathcal{P}(s|d) = &\left| 2 \pi \left(S^{-1} + M   \right)^{-1}\right|^{-\frac{1}{2}}\, \mathrm{e}^{ -\frac{1}{2} \left(s - j\right)^\dagger \left( S^{-1} + M  \right) \left(s - j\right)}.
 \end{split}
\end{equation}

We want to find a diagonal approximation $\tilde{M}$ to the measurement precision. The information loss by replacing $M$ with $\tilde{M}$ is to be minimal. It is quantified by the Kullback-Leibler divergence,
\begin{equation}
\begin{split}
 D_\mathrm{KL}\!\!\left[ \tilde{\mathcal{P}}(s|d) | \mathcal{P}(s|d) \right] = \int\!\!\mathcal{D}s \ \tilde{\mathcal{P}}(s|d)\, \ln\frac{\tilde{\mathcal{P}}(s|d)}{\mathcal{P}(s|d)}. \\
\end{split}
\end{equation}
For now we also introduce an approximate $\tilde{\jmath}$ to see if this implies corrections to $\tilde{M}$. We will equate it with $j$ later.
Dropping all terms independent of $\tilde{M}$ (denoted by $\hat{=}$) the Kullback-Leibler divergence is
\begin{equation} \begin{split}
D_{KL} \ \hat{=} & \ \frac{1}{2}\mathrm{tr}\!\left[R^{\dagger}N^{-1}R\left(D\tilde{\jmath}\tilde{\jmath}^{\dagger}D+D\right)\right]-\frac{1}{2}dN^{-1}RD\tilde{\jmath} \\
& - \frac{1}{2}\tilde{\jmath}^{\dagger}DR^{\dagger}N^{-1}d-\frac{1}{2}\mathrm{tr}\!\left[\tilde{M}\left(D\tilde{\jmath}\tilde{\jmath}^{\dagger}D+D\right)\right]\\
 & +\frac{1}{2}\tilde{\jmath}^{\dagger}D\tilde{\jmath}+\frac{1}{2}\mathrm{tr}\!\left[\ln\ D^{-1}\right],
\end{split} \end{equation} 
where $D = \left( S^{-1} + \tilde{M} \right)^{-1}$.
By setting $S^{-1}\rightarrow0$ (corresponding to a flat prior) we remove the influence of the prior. Since $\tilde{M}$ is chosen to be diagonal, $\tilde{M}_{kq}=\delta(k-q)\,g_{q}$, we have
$D^{-1}=\delta(k-q)\,g_{q}$ and $D=\delta(k-q)\,g_{q}^{-1}$.
This simplifies the Kullback-Leibler divergence further,
\begin{equation} \begin{split}
D_{KL} \ \hat{=} &\ \frac{1}{2}\int\!\! \mathrm{d}q\left[\left(R^{\dagger}N^{-1}R\right)_{qq}\left(|\tilde{\jmath}_{q}^{2}|/g_{q}^{2}+g_{q}^{-1}\right)\right]\\
& -\int\!\! \mathrm{d}q\,\mathfrak{R}\!\left[\left(R^{\dagger}N^{-1}d\right)_{q}^{*}\left(\tilde{\jmath}_{q}/g_{q}\right)\right]\\
& -\frac{1}{2}\int\!\! \mathrm{d}q\left[g_{q}\left(|\tilde{\jmath}_{q}^{2}|/g_{q}^{2}+g_{q}^{-1}\right)\right]\\
& +\frac{1}{2}\int\!\! \mathrm{d}q\,\frac{|\tilde{\jmath}_{q}^{2}|}{g_{q}}+\frac{1}{2}\int\!\! \mathrm{d}q\, \ln g_{q},
\end{split} \end{equation} 
where $\mathfrak{R}\!\left(\cdot\right)$ denotes the real part of a complex value.
Its functional derivative with respect to $g$ is
\begin{equation} \begin{split}
\frac{\delta}{\delta g_{q}}D_{KL} = & -\frac{1}{2}\left[\left(R^{\dagger}N^{-1}R\right)_{qq}\left(2|\tilde{\jmath}_{q}^{2}|/g_{q}^{3}+g_{q}^{-2}\right)\right]\\
 & +\mathfrak{R}\!\left[\left(R^{\dagger}N^{-1}d\right)_{q}^{*}\left(\tilde{\jmath}_{q}/g_{q}^{2}\right)\right]\\
 & +\frac{1}{2}|\tilde{\jmath}_{q}^{2}|/g_{q}^{2}-\frac{1}{2}|\tilde{\jmath}_{q}^{2}|/g_{q}^{2}+\frac{1}{2g_{q}}.
\end{split} \end{equation} 
Setting this derivative to zero and solving for $g$ yields
\begin{equation} \begin{split}
g_{q} & = \frac{1}{2}\left(R^{\dagger}N^{-1}R\right)_{qq}-\mathfrak{R}\!\left[\left(R^{\dagger}N^{-1}d\right)_{q}^{*}\tilde{\jmath}_{q}\right]\\
& \pm\,\left\{\frac{1}{4}\left(R^{\dagger}N^{-1}R\right)_{qq}^{2}+\mathfrak{R}\!\left[\left(R^{\dagger}N^{-1}d\right)_{q}^{*}\tilde{\jmath}_{q}\right]^{2}\right. \\
 & \left.-\left(R^{\dagger}N^{-1}R\right)_{qq}\mathfrak{R}\!\left[\left(R^{\dagger}N^{-1}d\right)_{q}^{*}\tilde{\jmath}_{q}\right]+2\left(R^{\dagger}N^{-1}R\right)_{qq}|\tilde{\jmath}_{q}^{2}| \right\}^{\frac{1}{2}}.
\end{split} \end{equation} 
Finally, by setting\footnote{
We could also minimize $D_\mathrm{KL}$ with respect to $\tilde{\jmath}$. This would yield
$\tilde{\jmath} = \tilde{M} M^{-1} j$. In our tests the inversion of $M$ was numerically problematic and we achieved better results using $\tilde{\jmath} = j$.
} $\tilde{\jmath}=j=R^\dagger N^{-1} d$ we arrive at
\begin{equation}
 g_{q} = \frac{1}{2}\left(R^{\dagger}N^{-1}R\right)_{qq}-|j_{q}^{2}|\pm\left[\frac{1}{2}\left(R^{\dagger}N^{-1}R\right)_{qq}+|j_{q}^{2}|\right]
\end{equation}
where the $+$ corresponds to minimal information loss.

The final result of this derivation is therefore that the measurement operator of Gaussian likelihood with linear response can be approximated by its diagonal,
\begin{equation}
 \tilde{M}_{kq} = \delta(k-q)\, \left( R^\dagger N^{-1} R \right)_{qq}.
\end{equation}
For the likelihood in this paper we choose to approximate the measurement precision operator after factoring out the primary beam\footnote{This corresponds to interpreting the primary beam as part of the signal for the sake of the derivation in this Appendix.}, thus setting 
\begin{equation}
 \tilde{F}(k,q; k',q') = \delta(k-k')\,\delta(q-q')\,F(k,q;k,q).
\end{equation}

\section{Relation to standard imaging practices}
\label{sec:relation_to_standard}

\corr{
The approximation procedure used in fast\textsc{Resolve} can be related to current practices in radio astronomical imaging.
The gridding of visibility data is standard in all imaging packages, since it is needed to perform the Fast Fourier Transform algorithm. Indeed, this procedure is also used in fast\textsc{Resolve}, where the response function $R$ from \textsc{Resolve} incorporates such a step using a Kaiser-Bessel gridding kernel \citep[see e.g.][]{Beatty-2005}. The present algorithm defines an approximation of this full gridding operation (see Sec.~\ref{sec:Mdiag} and Appendix~\ref{sec:DKL_minimization}), which is optimal in an information theoretical sense for the specific choice of likelihood used in \textsc{Resolve}. An analogy for this procedure from classical imaging would be any attempt to run CLEAN as a gridded imager, solely on the gridded UV-data, either in its original form known as H{\"o}gbom-CLEAN \citep{Hoegbom-1974} or even starting with an initial dirty image only derived from gridded visibilities. The latter was explored to reduce the needed disk space, memory size and computing time, indeed similar to the case of fast\textsc{Resolve}.
}

\corr{
In real standard practice, this approach was not as successful as the so-called major-minor-cycle framework \citep{Clark-1980, Schwab-1984}, which is nowadays regularly used in imaging packages using CLEAN. Here, the algorithm performs a subset of its iterations in gridded image space (the minor cycle) to transform back at specific intervals into the full de-gridded visibility space to perform a chi-square minimization on the full data with the current CLEAN model. One could use fast\textsc{Resolve} in a similar manner as a minor-cycle-like method, intersecting runs of the full \textsc{Resolve} algorithm as a major cycle step between the iterations of fast\textsc{Resolve} to increase the fidelity of the final image reconstruction. However, for the presented applications in this paper, we have refrained from exploring this option. Considering the simulated runs in Appendix~\ref{sec:mock_diffuse}, it might actually improve the reconstruction further.
}
 
\section{Approximate propagator}
\label{sec:approximate-propagator}

The filter equation \eqref{eq:MAP-equations-p} involves the evaluation of the propagator $D$. It is defined implicitly via its inverse \citep[][Eq.~(28)]{Junklewitz-2016},
\begin{equation}
\begin{split}
 D^{-1}(\vec{x},\vec{x}') = &\ S^{-1}(\vec{x}, \vec{x}') + e^{m(\vec{x})}\, M(\vec{x},\vec{x}')\, e^{m(\vec{x}')}\\
 & + \delta(\vec{x}-\vec{x}') e^{m(\vec{x})} \left(  M e^m - j \right)\!(\vec{x}).
 \label{eq:real_propagator}
 \end{split}
\end{equation}
Calculating the trace of $D$ is numerically expensive since a numerical inversion method like the conjugate gradient method has to be applied several times. The conjugate gradient method is efficient in inverting symmetric positive definite operators. Mathematically, the second derivative has to fulfill both conditions at the minimum. In numerical practice positive definiteness can be violated by the $\left(  M e^m - j \right)$ term. This term is proportional to the difference between real data and reconstucted data. Setting it to zero is therefore a reasonable approximation that ensures positive definiteness,
\begin{equation}
\begin{split}
 D^{-1}(\vec{x},\vec{x}') \approx &\ S^{-1}(\vec{x}, \vec{x}') + e^{m(\vec{x})}\, M(\vec{x},\vec{x}')\, e^{m(\vec{x}')}.\\
 \label{eq:approx_propagator}
 \end{split}
\end{equation}
This form of $D$ is used to calculate the uncertainty map described in Sec.~\ref{sec:uncertainty}.

For the power spectrum estimation in Eq.~\eqref{eq:MAP-equations-p} we can approximate $D$ even further to avoid the expensive numerical inversion altogether:
\begin{equation}
 D^{-1}(\vec{x},\vec{x}') \approx \ S^{-1}(\vec{x}, \vec{x}') + \left\langle \hat{B}e^{m} \right\rangle_{(\vec{x})}\, \tilde{F}(\vec{x},\vec{x}')\, \left\langle \hat{B}e^{m} \right\rangle_{(\vec{x})},
 \label{eq:approx_propagator2}
\end{equation}
with $\left\langle \cdot \right\rangle_{(\vec{x})}$ denoting the spatial average,
\begin{equation}
 \left\langle \hat{B}e^{m} \right\rangle_{\vec{x}} := \left.\int\mathrm{d}\vec{x}\, B(\vec{x}) e^{m(\vec{x})} \right/\int\mathrm{d}\vec{x}\, 1.
\end{equation}
This is of course a strong approximation to Eq.~\eqref{eq:real_propagator}, but in numerical tests Eq.~\eqref{eq:MAP-equations-p} still provided the desired fixed point. The advantage is that since both, $S$ and $\tilde{F}$ are diagonal in Fourier space, $D^{-1}$ is diagonal as well and the inversion is trivial.

\section{Likelihood variance estimation}
\label{sec:variance-derivation}

In this section we show how the variances of a Gaussian likelihood can be estimated. To that end we follow the derivation of \cite{Oppermann-2011}.
We assume that the noise is signal independent and that the diagonal basis of the noise covariance matrix is known,
\begin{equation}
\begin{split}
 d & = R(s) + n,\\
 \mathcal{P}(n|s) & = \mathcal{P}(n) = \mathcal{G}(n,N),\\
 N_{ij} & = \delta_{ij} \sigma^2_i.
\end{split}
\end{equation}
The diagonal entries $\sigma^2_i$ are unknown. Since they are variances, they have to be positive.
Therefore, a convenient choice for their prior is an inverse Gamma distribution,
\begin{equation}
 \mathcal{P}(\sigma^2) = \prod\limits_i \frac{1}{r_i\, \Gamma\!\left( \beta_i - 1 \right)} \left( \frac{\sigma^2_i}{r_i} \right)^{-\beta_i} \exp\!\left( - \frac{r_i}{\sigma^2_i} \right).
\end{equation}
The inverse Gamma distribution is defined by two parameters, $r$ and $\beta$. We will discuss canonical choices for them later on. It is furthermore beneficial to parametrize $\sigma^2_i$ by its logarithm,
\begin{equation}
\begin{split}
 \eta_i :=&\ \ln(\sigma^2_i),\\
 \mathcal{P}(\eta) =&\ \left.\mathcal{P}(\sigma^2)\right|_{\sigma^2 = \mathrm{e}^\eta} \prod\limits_i \mathrm{e}^{\eta_i} 
\end{split}
\end{equation}
An estimate for $\eta$ can now be derived using a joint maximum a posteriori Ansatz, i.e. we maximize the joint probability of $\eta$ and $s$ given the data. The joint maximum yields the same estimate for $s$ as if $\eta$ was known and the estimate for $\sigma^2 \equiv \mathrm{e}^\eta$ is
\begin{equation}
 \sigma^2_i = \frac{r_i + \frac{1}{2}\left| \left(R(s) - d  \right)_i\right|^2}{\beta_i-\frac{1}{2}}.
\end{equation}
The prior for $\sigma^2$ can be made uninformative by choosing $r_i \rightarrow 0$ and $\beta_i \rightarrow 1$. In this limit the inverse Gamma prior becomes Jeffrey's prior, which is flat in $\eta$,
\begin{equation}
 \left.\mathcal{P}(\eta)\right|_{r=0,\,\beta=1} = \mathrm{const.}
\end{equation}
Under this choice the estimate for $\sigma^2$ simplifies further to
\begin{equation}
  \sigma^2_i = \left| \left(R(s) - d  \right)_i\right|^2.
\end{equation}
This means that the $\sigma^2$ are chosen to set the $\chi^2$ value of each data point to 1.
To avoid the singularity at $\left(R(s) - d\right)_i=0$, we regularize $\sigma^2$ by
\begin{equation}
 \sigma^2_i \rightarrow t\, \sigma^2_i + (1-t) \frac{1}{N_\mathrm{data}} \sum_j \sigma^2_j,
\end{equation}
with $t \in [ 0, 1 ]$. In practice we choose $0$ for well-calibrated data sets and $t=0.9$ otherwise.

\section{Point source procedure}
\label{sec:point-source-details}

The point source contribution $I_\mathrm{point}$ is determined by iteratively collecting the brightest pixels of the maximum a posteriori map and subtracting their contribution from the data. The iterative procedure is stopped when the amount of pixels containing a point source reaches a cut-off criterion,
\begin{equation}
 \sum\limits_{\vec{x}} \Theta(I_\mathrm{point}(\vec{x})>0) > N_\mathrm{point},
 \label{eq:cut-off}
\end{equation}
where $\Theta$ is the indicator function,
\begin{equation}
 \Theta(I_\mathrm{point}(\vec{x})>0) = \begin{cases}
                                        1 \quad & \mathrm{if}\ \ I_\mathrm{point}(\vec{x})>0\\
                                        0 & \mathrm{else}.
                                       \end{cases}
\end{equation}
The cut-off criterion $N_\mathrm{point}$ should be chosen according to the expected amount of point sources in the image and the width of the point spread function. We are not aware of a canonical way to choose it.

Assuming a power spectrum $p$ and noise variances $\sigma_i^2$ start with an initial $I_\mathrm{point}$ (e.g.~0):
\begin{enumerate}
 \item calculate an estimate for $I = e^m$ by solving Eq.~\eqref{eq:MAP-equations-m} \label{iter:MAP}
 \item identify the maximal intensity $I_\mathrm{max} = \underset{\vec{x}}{\mathrm{max}}\,e^{m(\vec{x})}$
 \item point sources locations are $\{\vec{x}\ |\ I(\vec{x}) > I_\mathrm{max}/t_\mathrm{point} \}$
 \item at every point source location $I_\mathrm{point}$ is increased by $I(\vec{x})\times(1-1/t)$ \label{iter:point_increase}
 \item if the cut-off criterion Eq.~\eqref{eq:cut-off} is reached undo point \ref{iter:point_increase} and skip point \ref{iter:data_decrease}
 \item change the data to $d \rightarrow d - R(I_\mathrm{point})$ and repeat from point \ref{iter:MAP} \label{iter:data_decrease}
\end{enumerate}

The regularization factor $t_\mathrm{point}$ has to be larger than $1$, but should ideally be small. A good compromise is $1.75$ in our experience.

\section{Simulation tests}
\label{sec:mocktest}

\subsection{Diffuse emission}
\label{sec:mock_diffuse}

\begin{figure*}
\centering
     \begin{tabular}{c c c}
     \begin{overpic}[width=0.31\textwidth, trim=0 0 0 0, clip]{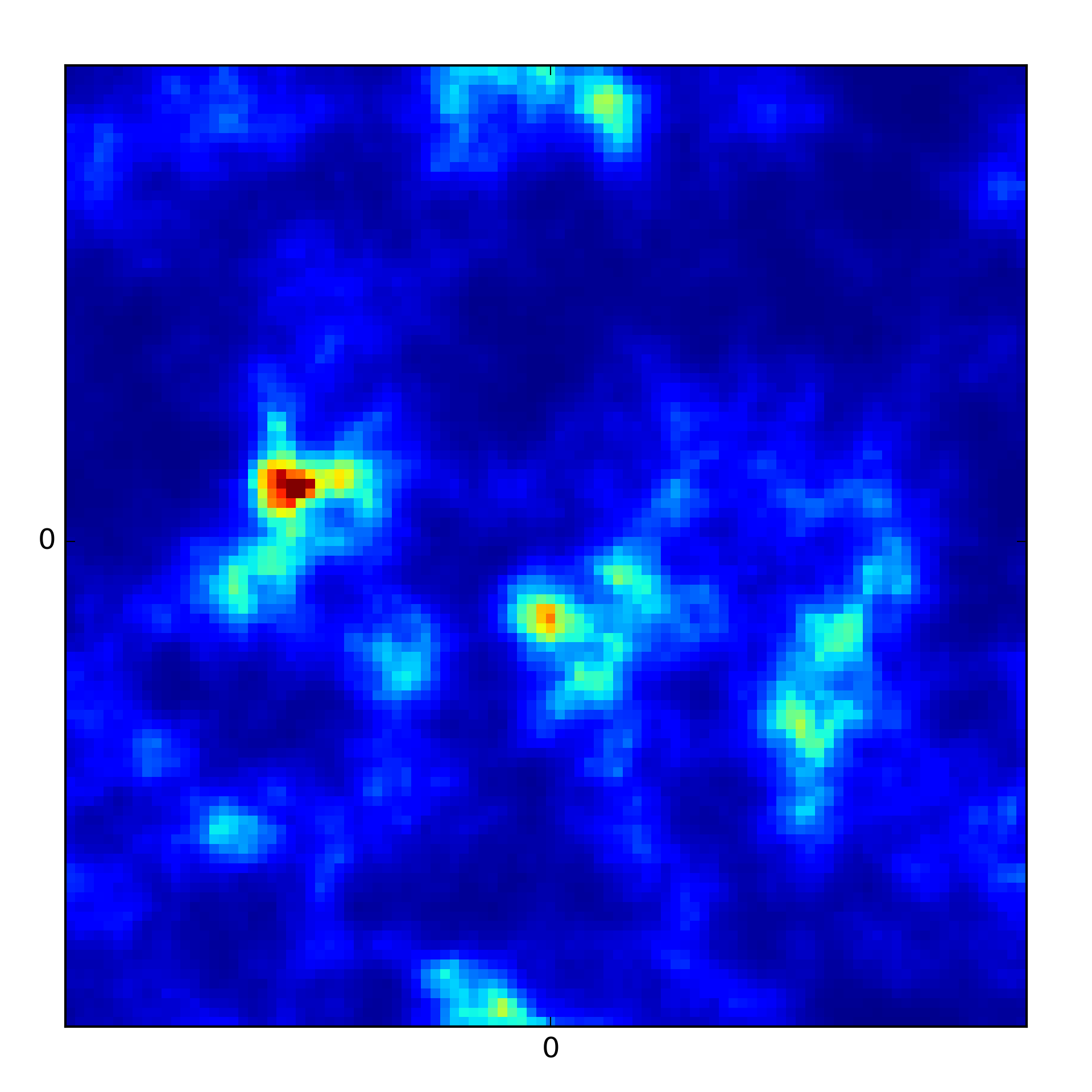}
      \put(40,80){\color{white}\textbf{original}}
     \end{overpic}
 & 
      \begin{overpic}[width=0.31\textwidth, trim=0 0 0 0, clip]{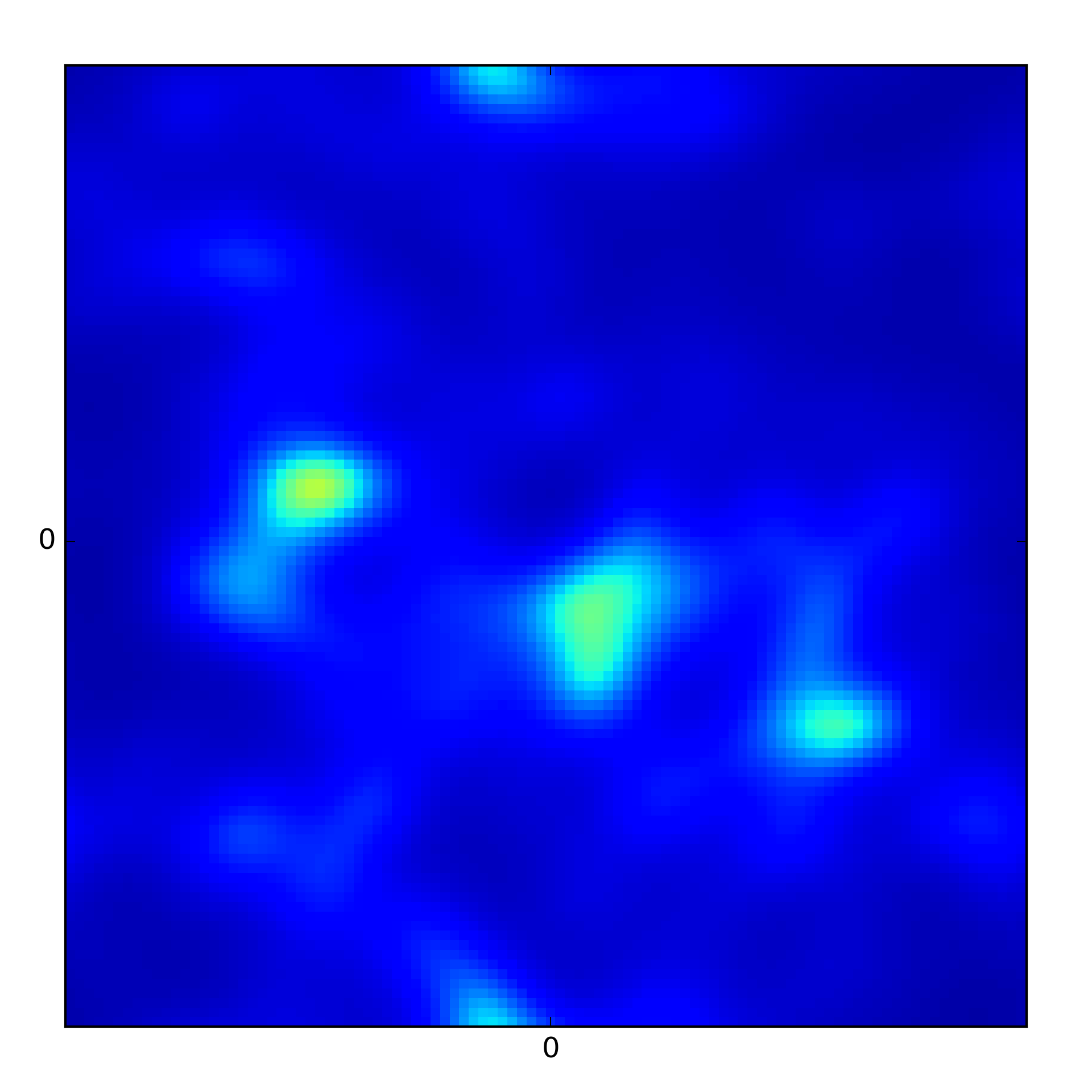}
      \put(25,80){\color{white}\textbf{\textsc{Resolve} SNR 0.1}}
     \end{overpic}
 & 
      \begin{overpic}[width=0.31\textwidth, trim=0 0 0 0, clip]{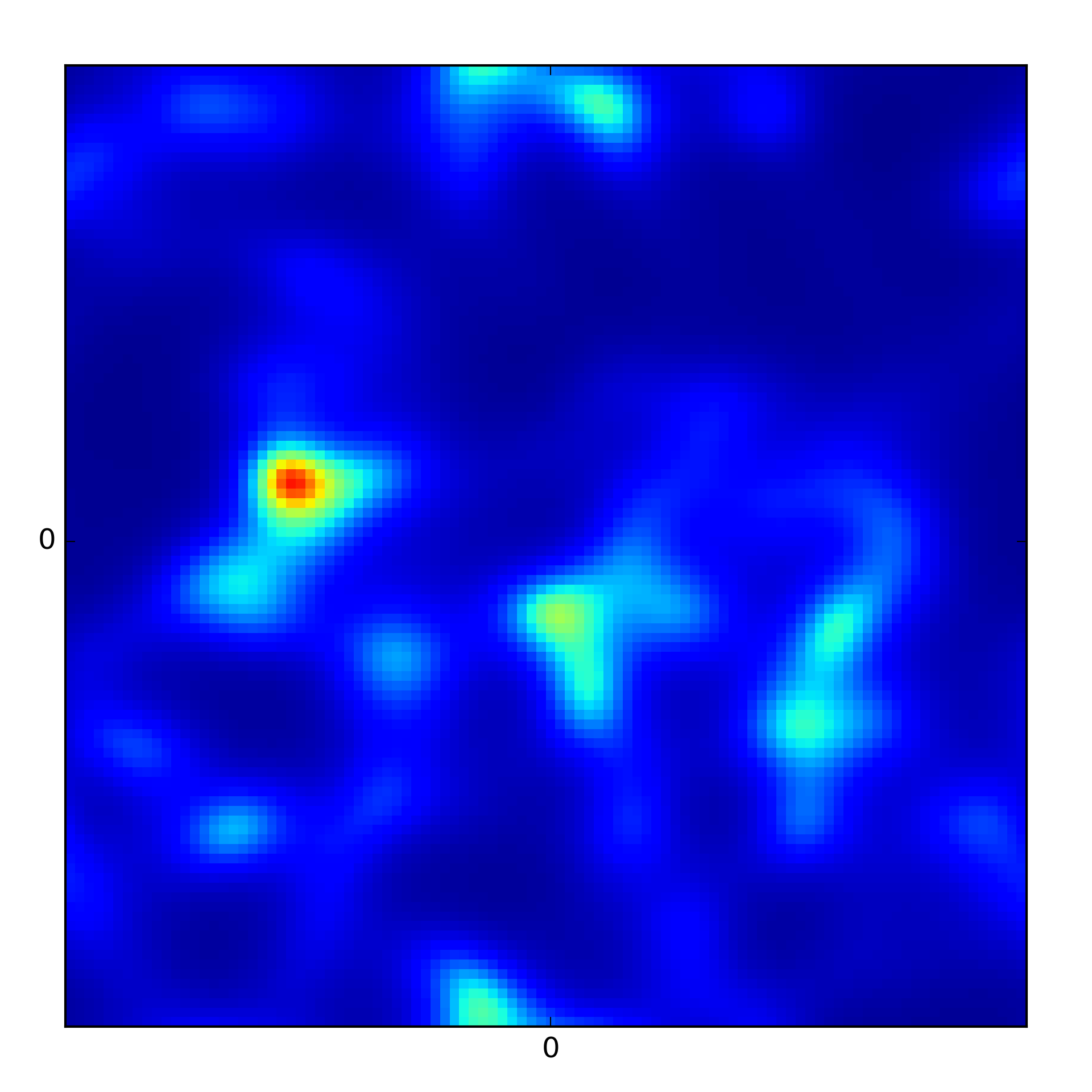}
      \put(25,80){\color{white}\textbf{\textsc{Resolve} SNR 20}}
     \end{overpic}
 \\
     \begin{overpic}[width=0.31\textwidth, trim=30 25 20 25, clip]{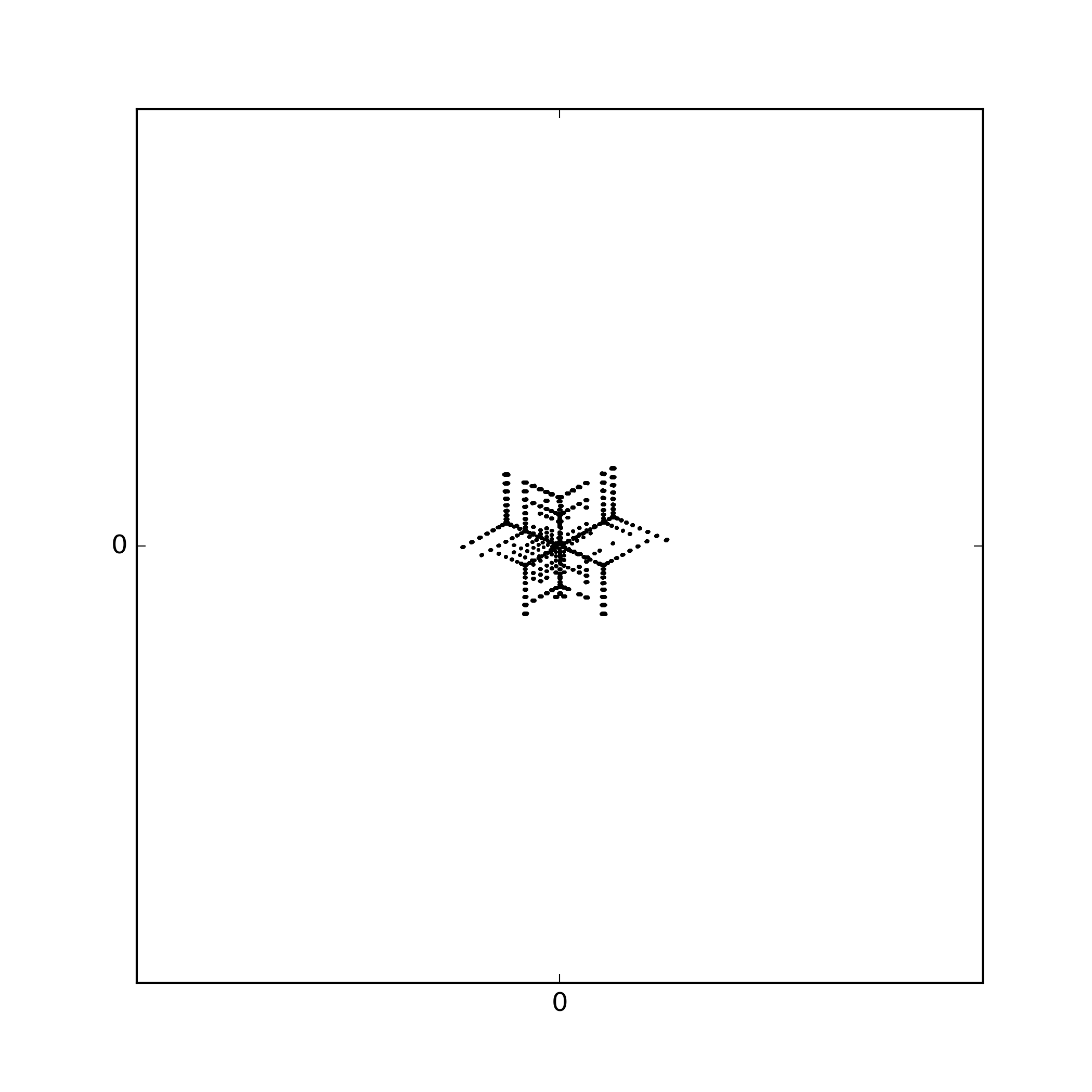}
      \put(35,80){\color{black}\textbf{uv coverage}}
     \end{overpic}
 &     
     \begin{overpic}[width=0.31\textwidth, trim=0 0 0 0, clip]{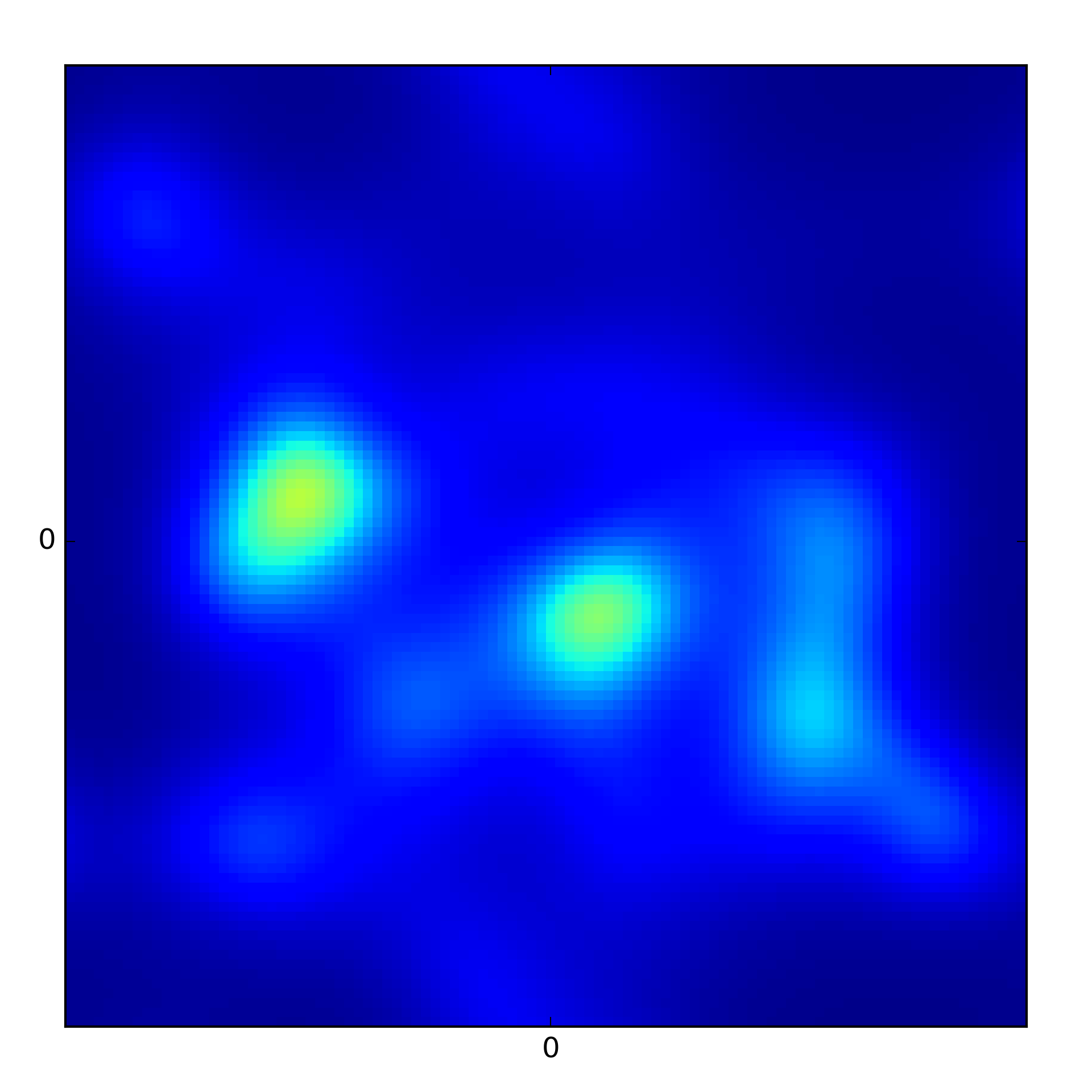}
      \put(20,80){\color{white}\textbf{fast\textsc{Resolve} SNR 0.1}}
     \end{overpic}
 &    
     \begin{overpic}[width=0.31\textwidth, trim=0 0 0 0, clip]{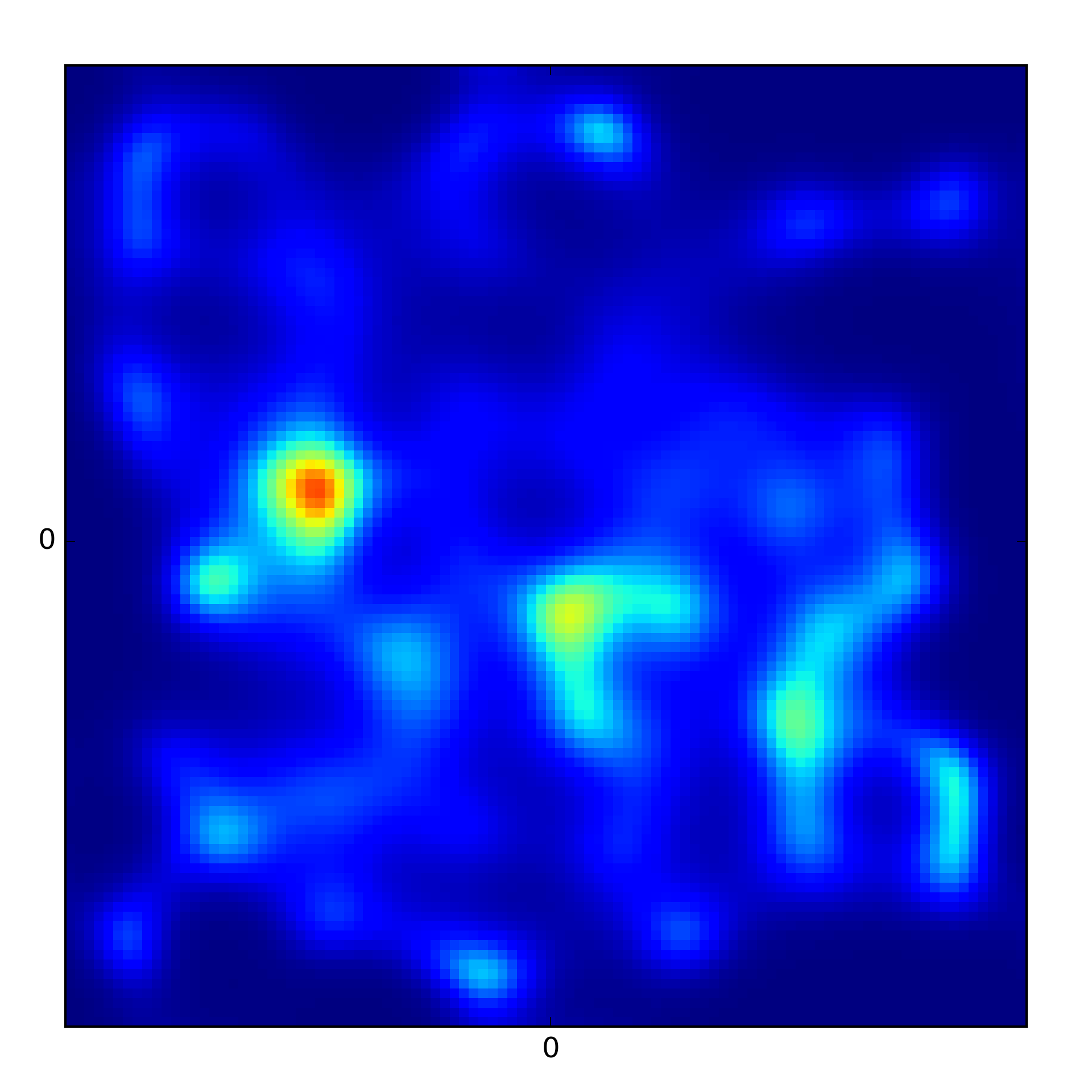}
      \put(20,80){\color{white}\textbf{fast\textsc{Resolve} SNR 20}}
     \end{overpic}
 \\
     \end{tabular}
 \caption{Simulation results for purely diffuse emission. All panels apart from the uv coverage use the same color scheme. The panel size of uv coverage corresponds to the Fourier scales populated by the mock signal.}
 \label{fig:simulations_diffuse}
\end{figure*}

In this section we validate the behavior of fast\textsc{Resolve} using simulated signal and data.
The simulated signal is the same as in the original publication of \textsc{Resolve} \citep[][Sec.~3]{Junklewitz-2016}. This allows for comparing the performance of these two \textsc{Resolve} variants.
As can be seen in Fig.~\ref{fig:simulations_diffuse} the results using fast\textsc{Resolve} are not quite as good as the ones using \textsc{Resolve}, but recover the most significant structure in a comparable quality.
This is due to the approximations made in the derivation of fast\textsc{Resolve}. However, fast\textsc{Resolve} detected automatically that there were no point sources present.
Furthermore, \textsc{Resolve} was provided with the measurement uncertainties while fast\textsc{Resolve} had to reconstruct them from the data. We can report that the reconstructed measurement uncertainty is mostly overestimated and reaches an upper limit even if the data are noise-free (using $t=0$). This leads to a more conservative final image.

\begin{figure*}
\centering
     \begin{tabular}{c c c}
     \begin{overpic}[width=0.31\textwidth, trim=0 0 0 0, clip]{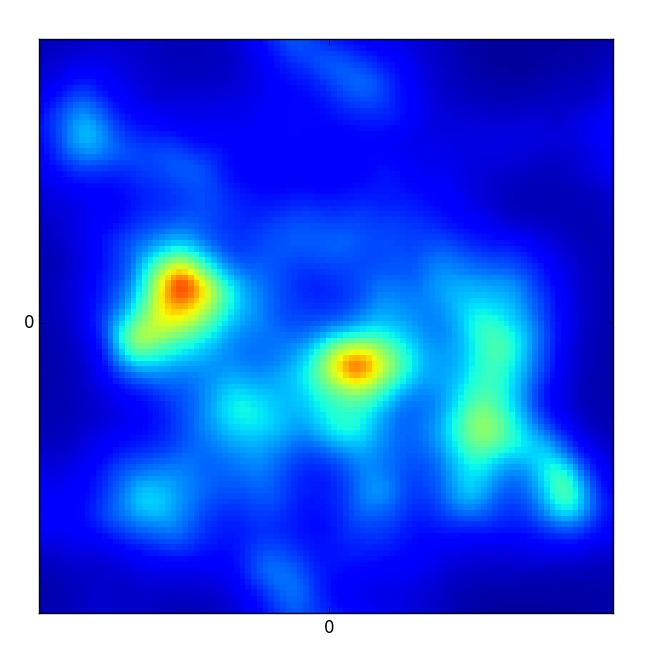}
      \put(47,80){{\color{white}$\boldsymbol{2\sigma}$}}
     \end{overpic}
 & 
      \begin{overpic}[width=0.31\textwidth, trim=0 0 0 0, clip]{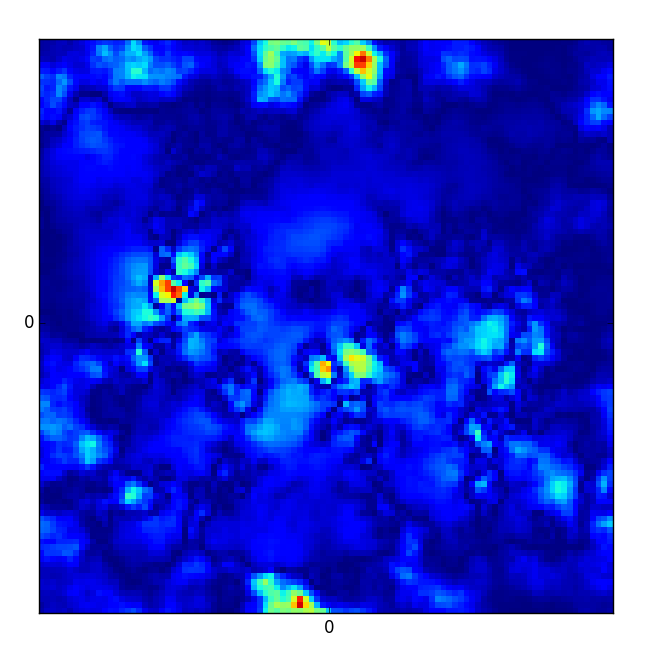}
      \put(30,80){\color{white}$\boldsymbol{|I_\mathrm{real} - I_\mathrm{recon}|}$}
     \end{overpic}
 \\

     \end{tabular}
 \caption{$2\sigma$ credibility interval (left panel) and absolute difference map (right panel). The color scheme is the same as in Fig.~\ref{fig:simulations_diffuse} up to a factor of two. This means a value of 1 in this figure has the same color as a value of 2 in Fig.~\ref{fig:simulations_diffuse}.}
 \label{fig:simulations_diffuse_unc}
\end{figure*}

\corr{
In Fig.~\ref{fig:simulations_diffuse_unc} we show the estimated uncertainty map and the absolute difference map for the low signal-to-noise case. Regions of large difference correspond mostly to regions of high uncertainty apart from the regions in the middle the top and bottom edge of the image. The absolute difference is smaller than the $1\sigma$ credibility interval for $55\%$ of the pixels and smaller than the $2\sigma$ interval for $85\%$ of the pixels. This confirms our suspicion that the approximative uncertainty map derived by a saddle point expansion of the posterior underestimates the real uncertainty. It is however of the right order of magnitude.
}

The \textsc{Resolve} runs used relatively loose convergence criteria and took roughly 1.9 hours each using two cores of an AMD Opteron 6376 @ 2.3Ghz.
The fast\textsc{Resolve} runs took 2.8 minutes each using relatively strict convergence criteria and running on a single core of an Intel Xeon E5-2650 v3 @ 2.3GHz. With loose convergence criteria, the run-time of fast\textsc{Resolve} dropped to 1.1 minutes. 
Even though it is not straight-forward to compare the covergence criteria of \text{Resolve} and fast\text{Resolve} (as mentioned in Sec.~\ref{sec:run-time}), this leads us to confirm the speed-up factor of roughly 100 we found in Sec.~\ref{sec:run-time}.

\subsection{point-like emission}

\begin{figure*}
\centering
     \begin{tabular}[t]{c c c}
     \begin{overpic}[width=0.31\textwidth, trim=0 0 0 0, clip]{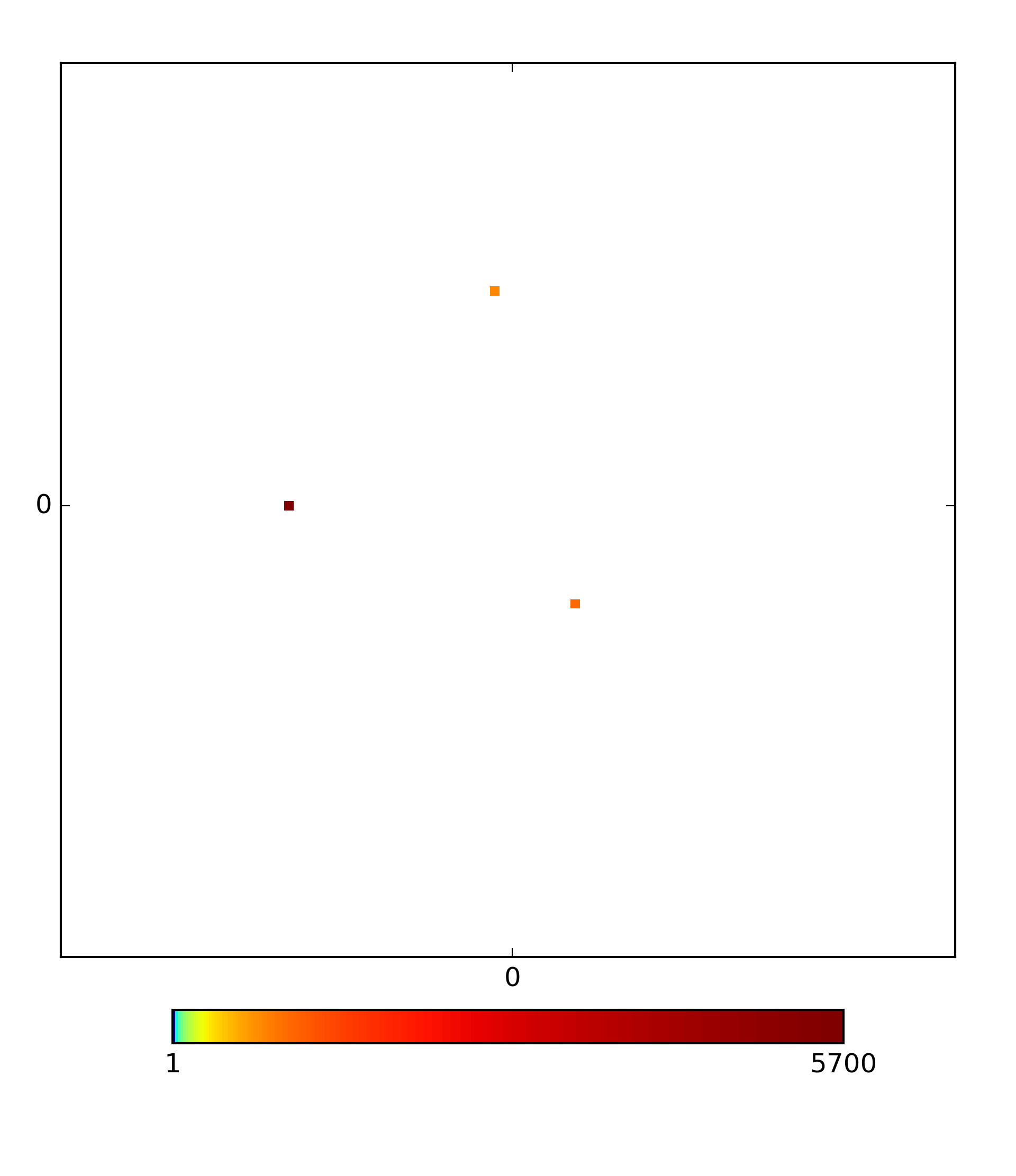}
      \put(20,80){\color{black}\textbf{point-like emission}}
     \end{overpic} 
 & 
     \raisebox{0.89cm}{
     \begin{overpic}[width=0.31\textwidth, trim=0 0 0 0, clip]{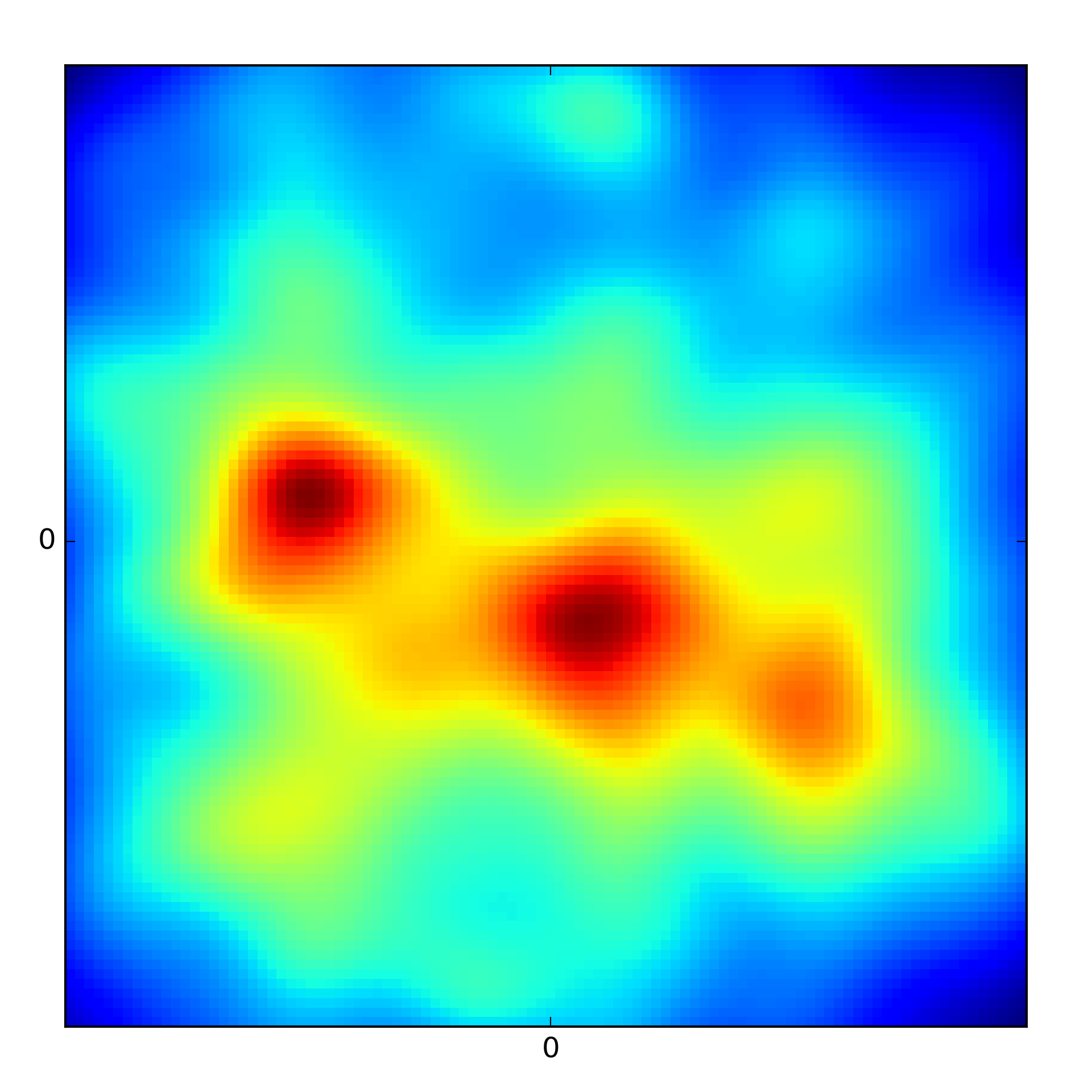}
      \put(12,80){\color{white}\textbf{dirty image -- diffuse only}}
     \end{overpic}
     }
 & 
      \begin{overpic}[width=0.31\textwidth, trim=0 0 0 0, clip]{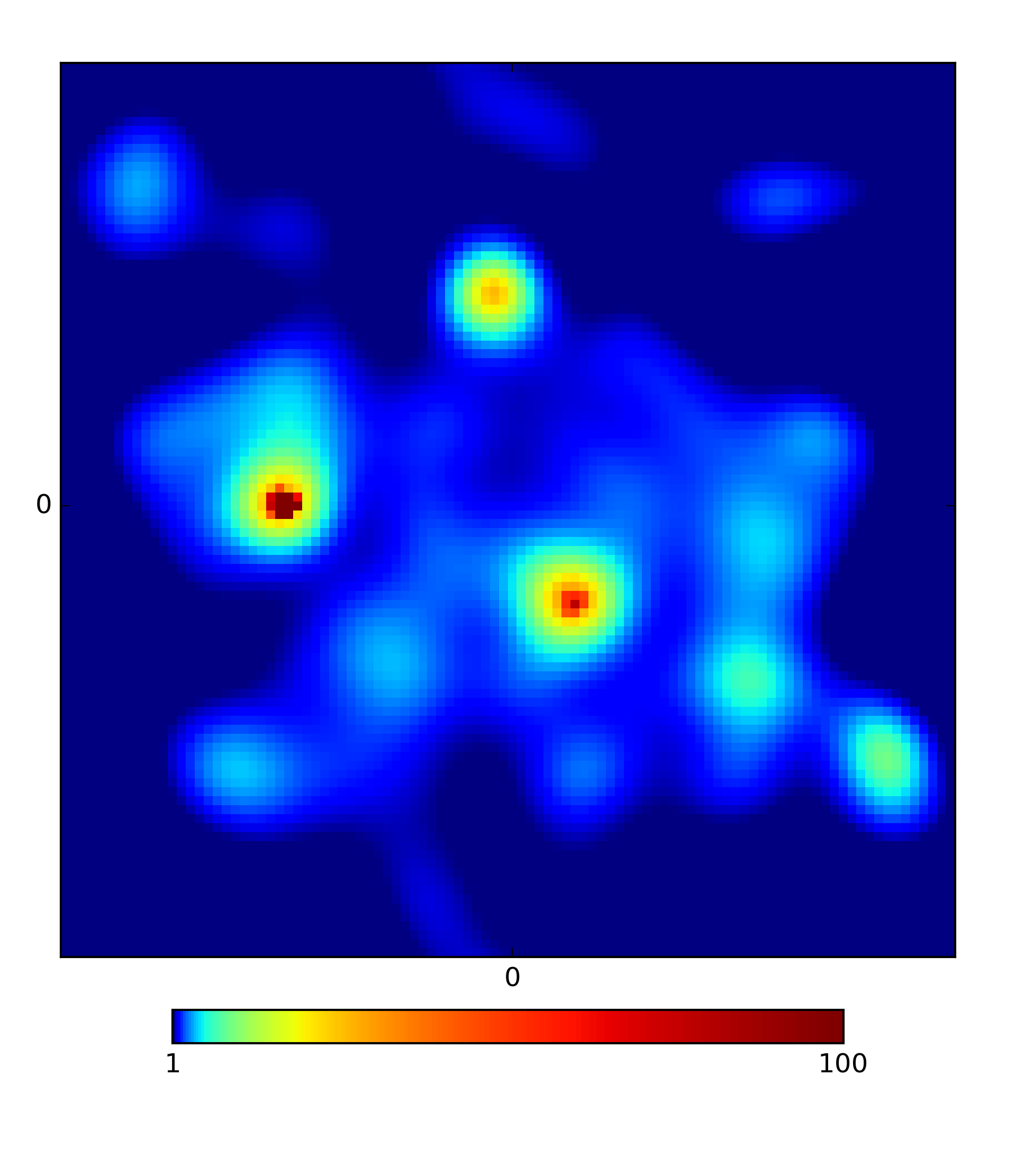}
      \put(34,80){\color{white}\textbf{SNR 0.1}}
     \end{overpic}
 \\
     \begin{overpic}[width=0.31\textwidth, trim=0 0 0 0, clip]{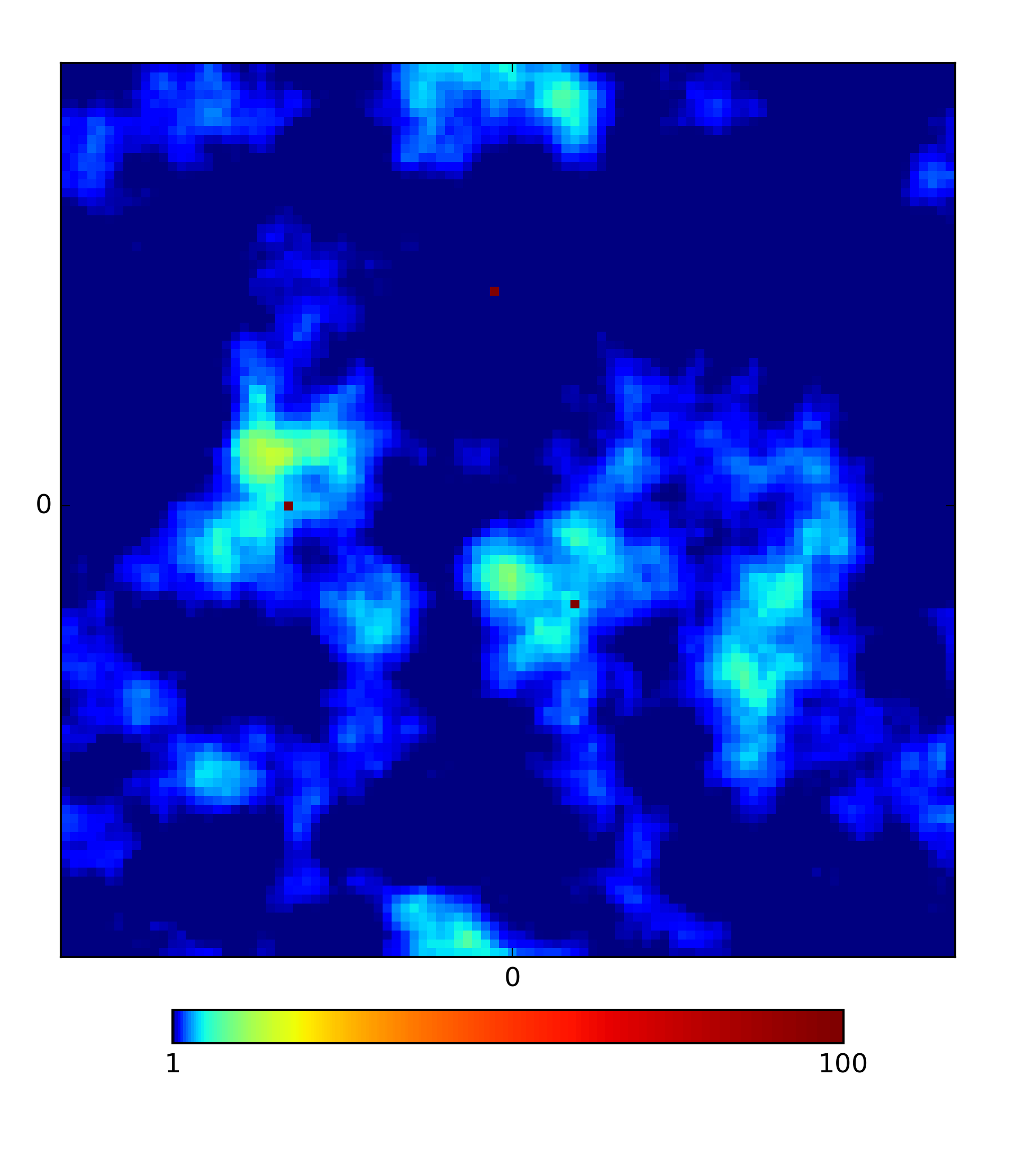}
      \put(25,80){\color{white}\textbf{total emission}}
     \end{overpic}
 & 
     \raisebox{0.89cm}{
     \begin{overpic}[width=0.31\textwidth, trim=0 0 0 0, clip]{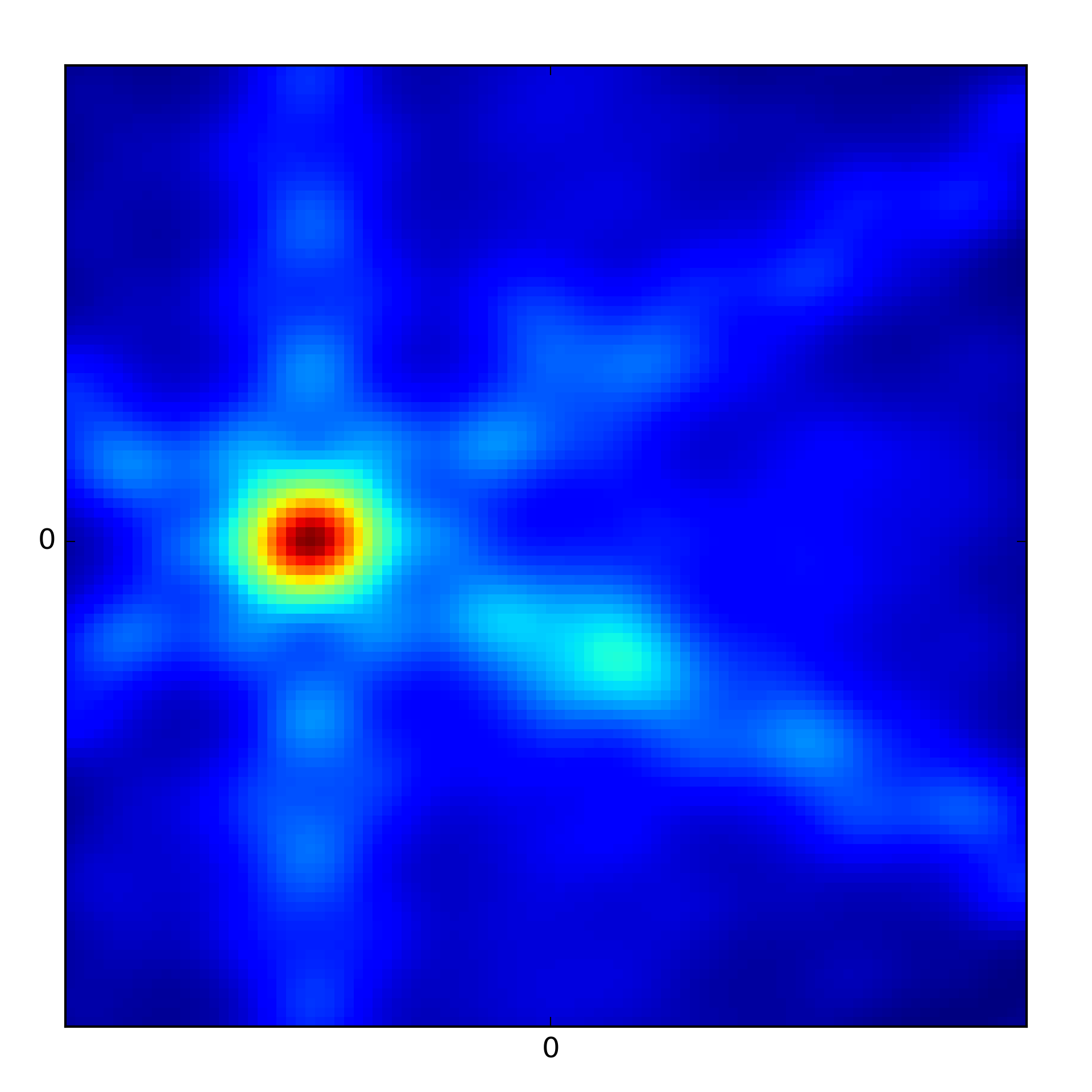}
      \put(33,80){\color{white}\textbf{dirty image}}
     \end{overpic}
     }
 & 
     \begin{overpic}[width=0.31\textwidth, trim=0 0 0 0, clip]{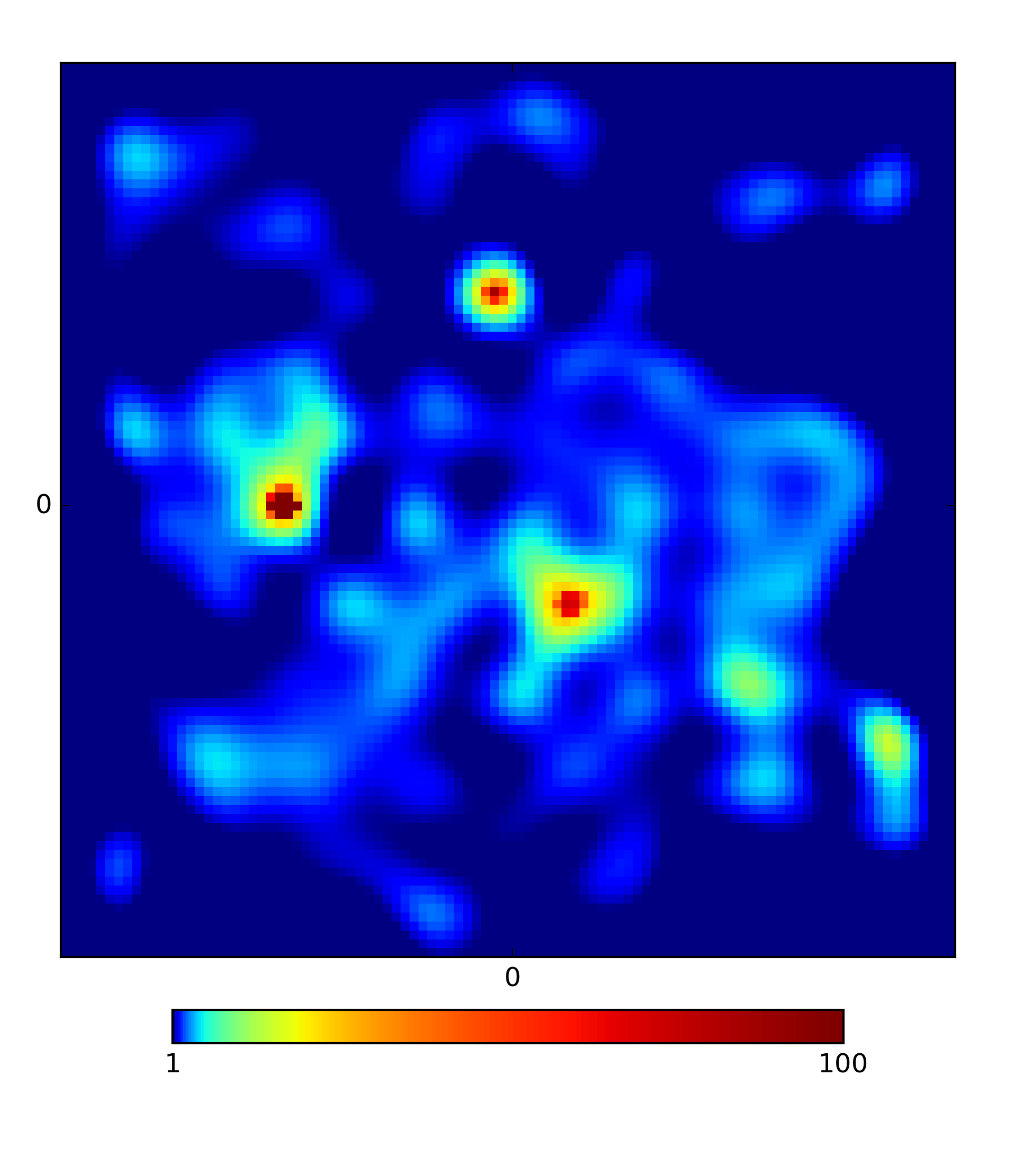}
      \put(34,80){\color{white}\textbf{SNR 20}}
     \end{overpic}
 \\
     \end{tabular}
 \caption{Simulation results for diffuse and point-like emission. The diffuse component and the uv-coverage were the same as in Fig.~\ref{fig:simulations_diffuse}.
 Left panels show the simulated emission, middle panels show the dirty images of the noiseless diffuse emisison (top) and noiseless total emission (bottom), right panels show the fast\textsc{Resolve} reconstructions.}
 \label{fig:simulations_point}
\end{figure*}

To test the performance of fast\textsc{Resolve} in the presence of point sources, we add three point sources to the simulated diffuse emission used in Appendix~\ref{sec:mock_diffuse}. The strongest peak of diffuse emission is at $14.9\mathrm{Jy}/\mathrm{px}$ in this test. The three point sources are at $5700\mathrm{Jy}/\mathrm{px}$, $1000\mathrm{Jy}/\mathrm{px}$, and $770\mathrm{Jy}/\mathrm{px}$ covering exactly one pixel at the cellsize of the simulation ($10^{-6} \mathrm{rad}$). The total flux of is $1.3\times 10^4 \mathrm{Jy}$ for the diffuse component and $7.5\times 10^3 \mathrm{Jy}$ for the point-like emission. In Fig.~\ref{fig:simulations_point} we show the simulated signal as well as the fast\textsc{Resolve} reconstructions at $S\!N\!R = 0.1$ and $S\!N\!R = 20$. The signal-to-noise value is with respect to the diffuse emission here. We use exactly the same noise contributions as in Appendix~\ref{sec:mock_diffuse}. As can be seen from the figure, the point-like emission is smoothed due to the point-spread-function, but some deconvolution is performed depending on the signal-to-noise ratio. The structure of the diffuse emission can be recovered in a comparable quality compared to the purely diffuse test in Appendix~\ref{sec:mock_diffuse} for this strength of point-like emission. We performed another simulation in which the brightness and total flux of the point-sources were 10 times greater (keeping all other parameters fixed). In that case only the most dominant features of the diffuse flux could be reconstructed. Most of it was obscured by the point sources. However, as long as the total flux of the point-like emission was smaller than or comparable to the diffuse flux, all our tests showed satisfactory results.

% \section{Possible $w$-term extension}
% \label{sec:w-term}
% 
% The interferometric equation (Eq.~\eqref{eq:interferometer}) is an approximation that is only valid for a small patch of the sky or a coplanar array of antennas. The full equation is
% \begin{equation}
%  V(u,v,w) = \int\!\!\mathrm{d}x\mathrm{d}y\,e^{-2\pi i \left( ux + vy + w\sqrt{1-x^2 - y^2}\right)}
%  \frac{B(x,y)\,I(x,y)}{\sqrt{1-x^2 - y^2}}.
%  \label{eq:uvw_equation}
% \end{equation}
% If $w$ vanishes (coplanar case) or $x$ and $y$ are small\footnote{In this case the phase factor $e^{-2\pi i w}$ can be transformed away.} (small patch) this equation goes to Eq.~\eqref{eq:interferometer}.
% 
% By redefining the beam $B(x,y)$ to a $w$-dependent (and complex) quantity $W(x,y,w)$, Eq.~\eqref{eq:uvw_equation} can be again brought into a form of a 2D Fourier transform,
% \begin{equation}
%  V(u,v,w) = \int\!\!\mathrm{d}x\mathrm{d}y\,e^{-2\pi i \left( ux + vy\right)}\,
%  W(x,y,w)\,I(x,y),
% \end{equation}
% where
% \begin{equation}
%  W(x,y,w) = \frac{B(x,y)\,e^{-2\pi i w\sqrt{1-x^2 - y^2}}}{\sqrt{1-x^2 - y^2}}.
% \end{equation}
% With this description the data set could be grouped into $l$ sets with equal $w$ with a different primary beam for each of them. This would mean $l$ Fourier transforms would be necessary to evaluate the likelihood. In reality this is of course impractical since $w$ would be just as irregularly spaced as $u$ and $v$ and consequently $m$ might be as large as the amount of data point $N_\mathrm{data}$.
%  

\section{Signal to noise}
\label{sec:SNR_formula}

In this paper we define the signal-to-noise ratio as
\begin{equation}
 S\!N\!R = \frac{1}{N_\mathrm{data}} \sum\limits_i \, \frac{\left| (R I)_i \right|^2}{\sigma_i^2}.
 \label{eq:SNR}
\end{equation}
There are other definitions of signal-to-noise, we chose this one for its simplicity. It can be estimated
by inserting the reconstructed $I$ or directly from the data as
\begin{equation}
 S\!N\!R \approx \frac{1}{N_\mathrm{data}} \sum\limits_i \, \frac{\left| d_i \right|^2}{\sigma_i^2} - 1.
\end{equation}
Due to the way fast\textsc{Resolve} estimates the likelihood variances $\sigma_i^2$ both estimates typically agree to a reasonable precision.
By inserting the corresponding components into Eq.~\eqref{eq:SNR} we can provide $S\!N\!R$ estimates for the point-like and diffuse flux separately.

\end{appendix}

\bibliographystyle{aa}
\bibliography{UV_sources}
 
\end{document}